\documentclass[aps,pre,preprint,eqsecnum]{revtex4-1} 
\usepackage{graphicx}
\usepackage{amssymb}
\usepackage{amsfonts}
\usepackage{amsmath}
\usepackage{mathrsfs} 
\usepackage{color}
\usepackage{lineno}
\usepackage{hyperref}
\providecommand\bnabla{\boldsymbol{\nabla}}

\providecommand\bcdot{\boldsymbol{\cdot}}

\providecommand\br{\mathbf{r}}

\providecommand\bv{\mathbf{v}}
\providecommand\bx{\mathbf{x}}

\providecommand\bJ{\mathbf{J}}

\providecommand\bzero{\mathbf{0}}

\providecommand\bn{\mathbf{\hat{n}}}

\providecommand\tI{\mathsf{I}}

\providecommand\tE{\mathsf{E}}

\providecommand\unit{\boldsymbol{\hat{\imath}}}

\newcommand{\ub}[1]{^{({#1})}}

\begin{document}

\title{Asymptotic modeling of Helmholtz resonators \\ including thermoviscous effects}
\author{Rodolfo Brand\~ao}
\author{Ory Schnitzer}
\affiliation{Department of Mathematics, Imperial College London, London SW7 2AZ, UK}

\begin{abstract}
We systematically employ the method of matched asymptotic expansions to model Helmholtz resonators, with thermoviscous effects incorporated starting from first principles and with the lumped parameters characterizing the neck and cavity geometries precisely defined and provided explicitly for a wide range of geometries. With an eye towards modeling acoustic metasurfaces, we consider resonators embedded in a rigid surface, each resonator consisting of an arbitrarily shaped cavity connected to the external half-space by a small cylindrical neck. The bulk of the analysis is devoted to the problem where a single resonator is subjected to a normally incident plane wave; the model is then extended using ``Foldy's method'' to the case of multiple resonators subjected to an arbitrary incident field. As an illustration, we derive critical-coupling conditions for optimal and perfect absorption by a single resonator and a model metasurface, respectively.
\end{abstract}

\maketitle

\section{Introduction}

Determining the acoustical response of a Helmholtz resonator, namely a hollow cavity connected to an exterior region through a small neck, 
is a long-standing problem in theoretical acoustics \cite{helmholtz:1860,rayleigh:1871,rayleigh:1896,ingard:1953}. The
Helmholtz resonance refers specifically to the fundamental ``breathing'' mode of the cavity, which occurs at a wavelength many times larger than the cavity. This is in contrast to all higher-order modes, which are related to the formation of standing waves. 
Helmholtz resonators have long been employed in acoustic devices such as perforated panel absorbers \cite{maa:1998}, mufflers \cite{yasuda:2013} and acoustical dampers \cite{zhao:2015}.
In recent years, arrays of Helmholtz resonators have also been widely used to construct acoustic metamaterials and metasurfaces exhibiting special phenomena such as effective negative stiffness \cite{fang:2006}, subwavelength focusing \cite{lemoult:2011}, perfect absorption \cite{jimenez:2016}, exceptional points \cite{ding:2016} and rainbow trapping \cite{jimenez:2017}.

Traditionally, the Helmholtz resonator is modeled as a mass-spring system, where the slug of air oscillating in the neck acts as a lumped mass and adiabatic compression of the air within the cavity provides the spring action \cite{rayleigh:1871}. From this intuitive model, the Helmholtz resonance frequency can be readily derived as
\begin{equation}\label{dimresf}
\omega = c\sqrt{\frac{S}{H V}},
\end{equation}
where c is the sound speed, $H$ and $S$ are respectively the length and cross-sectional area of the neck and $V$ is the volume of the cavity. Let $\epsilon\ll1$ represent the ratio between the characteristic sizes of the neck and cavity such that $H= O\left(\epsilon V^{1/3}\right)$ and $S = O\left(\epsilon^2 V^{2/3}\right)$. Then it follows from  \eqref{dimresf} that $\omega V^{1/3}/c = O\left(\epsilon^{1/2}\right)$, showing that the resonance frequency lies in the subwavelength regime.

The mass-spring model of a Helmholtz resonator provides only a rough prediction for its resonance frequency. In particular, whereas the geometric quantities appearing in \eqref{dimresf} are defined ambiguously, it has been shown that the shapes of both the neck and cavity may appreciably affect the response of the resonator \cite{ingard:1950,ingard:1953,zinn:1970, alster:1972,panton:1975,chanaud:1994,hersh:2003}. 
Furthermore, the mass-spring model does not account for any form of damping and hence cannot be used by itself to describe the acoustical response of the resonator to external forcing. Clearly the response of any resonator is bounded and thence either radiation damping or dissipative losses,
or both, must become important at frequencies sufficiently close to resonance.  

Surprisingly, these deficiencies have never been fully resolved in the literature. Indeed, it remains the case that most analytical models of Helmholtz resonators involve heuristics, resulting in additional physical ambiguities and parameters that need to be verified on an \textit{ad hoc} basis \cite{ingard:1953,zinn:1970, alster:1972,panton:1975,howe:1976,chanaud:1994,sugimoto:1995,hersh:2003,komkin:2017,jimenez:2016}. There have been several asymptotic studies based on the small-neck limit $\epsilon\ll1$, using matched asymptotic expansions \cite{bigg:1982,monkewitz:1985,monkewitz2:1985} and layer-potential techniques \cite{mohring:1999,ammari:2015}. However, among other issues, the neglect of thermoviscous effects in those studies severely limits their applicability. %To address these deficiencies, the method of matched asymptotic expansions \cite{bigg:1982,monkewitz:1985,monkewitz2:1985} and layer-potential techniques \cite{mohring:1999,ammari:2015} have been employed towards deriving an asymptotic model of the Helmholtz resonator in the  small-neck limit $\epsilon\ll1$. The neglect of thermoviscous effects in all these analyses, however, severely limits their applicability.

In this paper we systematically employ the method of matched asymptotic expansions to model Helmholtz resonators in the small-neck limit, with thermoviscous effects incorporated starting from first principles. With an eye towards modeling acoustic metasurfaces, we consider the linear response of resonators embedded in a rigid surface, each resonator consisting of an arbitrarily shaped cavity connected to the external half-space by a small cylindrical neck. Thermoviscous effects are modeled assuming that the fluid is an ideal gas. We shall at first focus on the case of a single resonator excited by a normally incident plane wave. Then, using Foldy's method \cite{carstensen:1947,martin:2006} we shall generalize to an arbitrary distribution of surface-embedded resonators excited by an arbitrary incident field.

 The method of matched asymptotic expansions allows us to exploit the spatial nonuniformity of the small-neck limit. For frequencies on the order of the Helmholtz resonance \eqref{dimresf}, the small-neck limit also constitutes a long-wavelength limit. Thus, the fluid domain naturally separates into a small neck region, a larger cavity region and an even larger, wavelength-scale, external region. Furthermore, we shall find that given the resonant nature of the problem the small-neck limit is also nonuniform in frequency, even when frequency is confined to the subwavelength regime implied by  \eqref{dimresf}. This point appears to have been overlooked in preceding analyses based on the method of matched asymptotic expansions \cite{bigg:1982,monkewitz:1985,monkewitz2:1985}. In contrast, as part of our analysis we shall identify and separately analyze a hierarchy of distinguished frequency intervals converging to  the Helmholtz resonance.

As we shall see, this asymptotic approach possesses several key advantages. First, it allows precisely defining the lumped parameters emerging from the analysis in terms of ``canonical'' problems depending only on geometry. In particular, our analysis reveals a natural linkage between the parameters characterizing the neck geometry and the viscous acoustic impedance of the cylindrical neck, for which we have recently provided accurate numerical values and asymptotic formulas as a function of the neck aspect ratio \cite{brandao:20xx}. Second, it provides new physical insights that, \textit{inter alia}, facilitate a systematic treatment of thermoviscous effects. In that context, our interest lies in the regime where the resonance remains weakly damped. This condition naturally translates to one on the thickness of the thermoviscous boundary layers, relative to the resonator dimensions, which we shall determine as part of the analysis. Of particular interest is the regime where thermoviscous dissipation is comparable to radiation damping, thence dissipation is maximized \cite{ingard:1953}. As an illustration, we derive critical-coupling conditions for optimal and perfect absorption of a plane wave by a single resonator and by a model metasurface, respectively.

The paper is organized as follows. In the next section, we 
formulate the problem for a single resonator in the absence of thermoviscous effects. We then analyze this ``lossless'' problem in \S\ref{sec:leading}--\S\ref{sec:rdamp}. In \S\ref{sec:dissipation} we extend our analysis by including thermoviscous effects.
In \S\ref{sec:discussion} we recapitulate and discuss our model of a single resonator. In \S\ref{sec:mult} we generalize to the case of multiple resonators. Concluding remarks are given in \S\ref{sec:conclusions}.

\section{Problem formulation without dissipation}\label{sec:pf}
\begin{figure}[t!]
\includegraphics[scale=0.35]{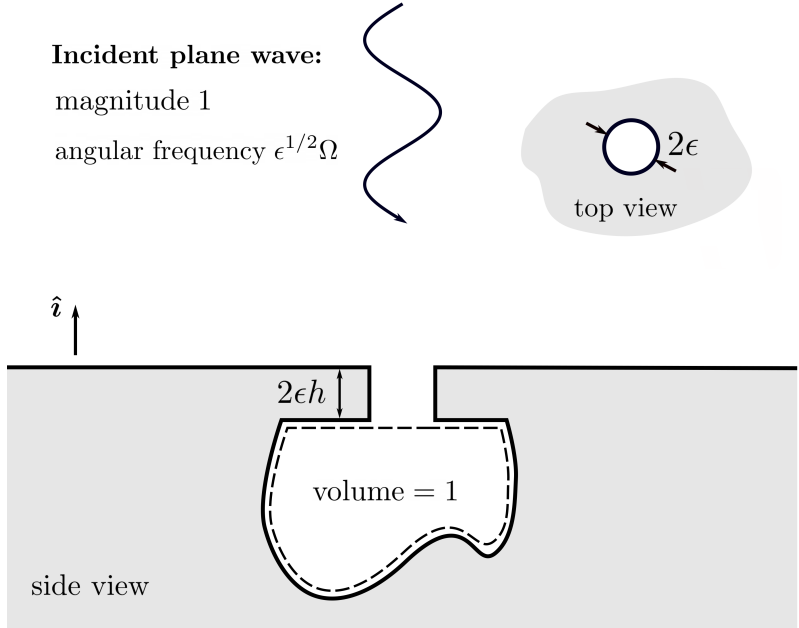}
\caption{Dimensionless schematic of a single Helmholtz resonator, formed of a cavity and a cylindrical neck, embedded in a rigid substrate and subjected to a normally incident plane wave.}
\label{fig:ske}
\end{figure}

Consider a Helmholtz resonator consisting of a cavity connected to an exterior half-space by a small neck, as illustrated in Fig.~\ref{fig:ske}. For simplicity, we assume that the surface is rigid, that the neck is a small cylindrical channel and that the cavity boundary is flat in some neighborhood of the neck opening. The volume of the cavity, excluding the neck, is denoted by $l^{3}$ and the radius and height of the neck are denoted by  $\epsilon l $ and $2 \epsilon h l $, respectively. For the time being, we neglect thermoviscous effects and assume that the resonator is forced by a normally incident pressure plane wave at fixed angular frequency $\omega$ and amplitude $p_{\infty}$.

Henceforth, we adopt a dimensionless convention where lengths are normalized by $l$ and $p$ denotes the  pressure normalized by $p_{\infty}$, with the factor $\exp(-i\omega t)$ suppressed in the usual way. We also define: (i) the unit vector $\unit$ co-linear with the cylinder axis and pointing towards the exterior domain; and (ii) three position vectors, $\bx$ and $\bx^{\pm}$, with $\bx$ measured from the center of the cylindrical neck and $\bx^{\pm}= \bx \mp \epsilon h \unit$. Our interest is in the fundamental (``Helmholtz'') resonance. Accordingly, \eqref{dimresf} suggests defining the dimensionless frequency 
\begin{equation}\label{Dimensionless frequency}
\Omega = \frac{\epsilon^{-1/2} \omega l}{c},
\end{equation}
where $c$ is the speed of sound. 

The field $p$ is governed by the Helmholtz equation
\begin{equation}\label{Helmholtz equation}
\nabla^2 p + \epsilon \Omega^2 p = 0
\end{equation}
in the fluid domain;
the Neumann condition 
\begin{equation}\label{Neumann condition}
\bn\bcdot\bnabla p = 0
\end{equation}
on the rigid boundary, where $\bn$ is the normal unit vector pointing into the fluid; and, at large distances from the resonator, a radiation condition imposed on $p-p\ub{i}$, wherein  
\begin{equation}\label{Incoming wave}
p\ub{i}=\exp\left(-i \epsilon^{1/2} \Omega\, \unit\bcdot \bx^+ \right)
\end{equation}
is the incident plane wave.

In the following sections \S\ref{sec:leading}--\S\ref{sec:rdamp} our aim is to study \eqref{Helmholtz equation}--\eqref{Incoming wave} in the limit $\epsilon\to0$. Since our interest is in the fundamental resonance of the cavity, we assume $\Omega=O(1)$, whereby \eqref{Dimensionless frequency} implies that the resonator is asymptotically small compared to the acoustic wavelength. Accordingly, the problem is characterized by three disparate length scales: the $O(\epsilon)$ neck radius and height, the $O(1)$ cavity size and the $O\left(\epsilon^{-1/2}\right)$ wavelength. (We  assume that $h$ is independent of $\epsilon$.) To exploit this separation of scales we shall employ the method of matched asymptotic expansions \cite{Van:pert,crighton:1992}. Owing to the resonant nature of the cavity's response, the small-neck limit is also nonuniform in frequency. Accordingly, we shall separately analyze several distinguished frequency regimes defined by their width about resonance.

In characterizing the acoustical response of the resonator, we shall focus on the pressure in the cavity and the diffracted field. It is well known that in the subwavelength regime the pressure is approximately uniform within the cavity; this will be confirmed by the analysis. As for the diffracted field, the exterior wave can be written, without approximation, 
in the form (see, e.g., \citep{crighton:1992}) 
\begin{equation}\label{farfie}
p = 2\cos\left(\epsilon^{1/2} \Omega\, \unit\bcdot \bx^+ \right) +  A\frac{\exp\left(i \epsilon^{1/2} \Omega |\mathbf{x}^+|\right)}{|\mathbf{x}^+|}  + \boldsymbol{B}\cdot \bnabla \frac{\exp\left(i \epsilon^{1/2} \Omega |\mathbf{x}^+|\right)}{|\mathbf{x}^+|} + \cdots.
\end{equation}
Here, the first term is the superposition of the incident wave \eqref{Incoming wave} and the plane wave reflected from the rigid wall. The second term represents a spherical wave diffracted by the resonator; it is proportional to the fundamental solution of the Helmholtz equation in three dimensions. Higher-order terms, formed of products of constant tensors and gradients of the latter fundamental solution, will be seen to be negligible at distances which are large compared to the neck radius. Accordingly, the diffracted field will be characterized by the complex-valued amplitude $A$.

\section{Off-resonance limit}\label{sec:leading}

Naively, our assumption that $\Omega=O(1)$ suggests studying the limit $\epsilon\to0$ with $\Omega$ held fixed. We shall refer to the approximation obtained in this limit as ``off-resonance,'' as we shall find that it breaks down as $\Omega$ approaches its value at resonance. The off-resonance analysis in this section formalizes the classical mass-spring model of the Helmholtz resonator and sets the stage for a systematic study of near-resonance frequencies in subsequent sections.

\subsection{Cavity region}
\label{sec:cavity off resonance}
\begin{figure}[t!]
\includegraphics[scale=0.4]{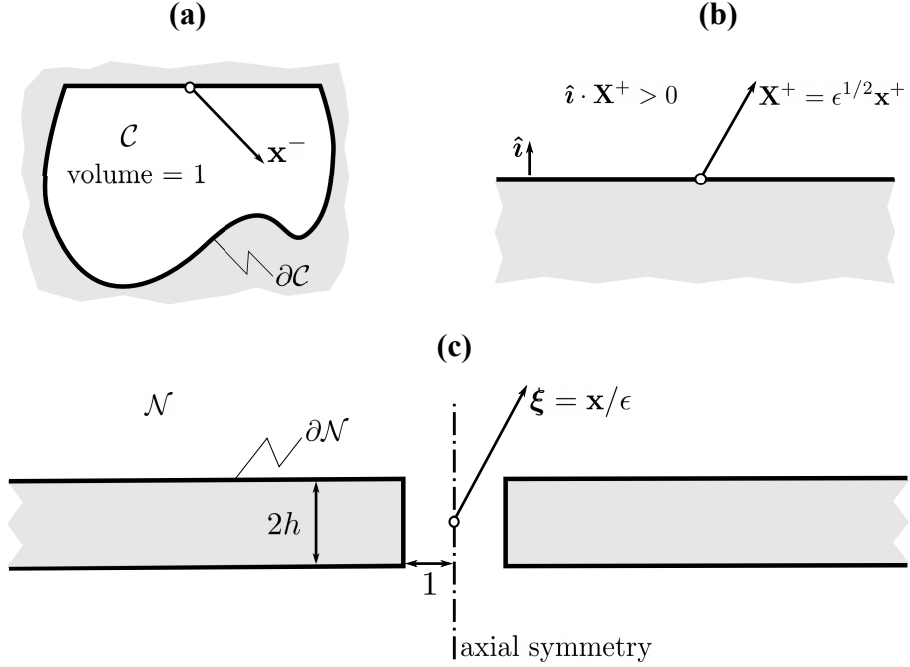}
\caption{Distinguished regions in the matched asymptotics analysis. (a) Cavity region. (b) Exterior region (wavelength scale). (c) Neck region. }
\label{domains}
\end{figure}

We begin by investigating the cavity region by holding $\bx^{-}$ fixed as $\epsilon\to0$. 
The limiting geometry is sketched in Fig.~\ref{domains}(a); it is identical to the exact cavity geometry, except that the neck opening degenerates to the point $\bx^{-} = \bf{0}$. The corresponding fluid domain is denoted by $\mathcal{C}$; note that the definition of this domain in terms of $\bx^{-}$ is independent of $\epsilon$.   We retain the symbol $p$ for the pressure field in the cavity region. It satisfies \eqref{Helmholtz equation} in $\mathcal{C}$ and the Neumann condition \eqref{Neumann condition} on $\partial{C}$ --- excluding the degenerate point $\bx^{-} = \bf{0}$, where the solution is subject to matching conditions and may be singular. 
Let us assume, subject to verification through matching, that the cavity pressure is comparable to the $O(1)$ incident field \eqref{Incoming wave}. 
Accordingly, we attempt an expansion in the form 
\begin{equation}\label{p off expansion}
p = p_0 + \epsilon p_{1} +\cdots \quad \text{as} \quad \epsilon\to0,
\end{equation}
with $\bx^{-}$ fixed, where the leading-order field $p_0$ satisfies Laplace's equation
\begin{equation}\label{p0 off equation}
\nabla^2 p_0 = 0 \quad \text{in} \quad \mathcal{C}
\end{equation}
and
\begin{equation}\label{p0 off condition}
\bn\bcdot\bnabla p_{0} = 0 \quad \text{on} \quad \partial\mathcal{C}.
\end{equation}
Since we expect the pressure within the resonator to be at least comparable to that in the neck, we can rule out the possibility of $p_0$ being singular at $\bx^-=\bzero$.  
It readily follows that $p_0$ is uniform, equal to a constant, say $\bar{p}_0$, to be determined.

From the $O(\epsilon)$ balances of \eqref{Helmholtz equation} and \eqref{Neumann condition}, the leading correction $p_1$ satisfies the forced Laplace's equation
\begin{equation}\label{p1 off equation}
\nabla^2 p_1 =- \Omega^2 \bar{p}_0 \quad \text{in} \quad \mathcal{C},
\end{equation}
with
\begin{equation}\label{p1 off neumann}
\bn\bcdot\bnabla p_{1} = 0 \quad \text{on} \quad \partial\mathcal{C} \setminus \{\bx^{-}=\bzero\}.
\end{equation}
Since neither $\Omega$ nor $\bar{p}_0$ vanish, application of the divergence theorem  to \eqref{p1 off equation} and \eqref{p1 off neumann} implies a monopole singularity at the degenerate point:
\begin{equation}\label{p1 off singularity}
p_1\sim -\frac{\Omega^2 \bar{p}_0}{2\pi |\bx^-|} + O(1) \quad \text{as} \quad |\bx^-|\to0.
\end{equation}
Dipole and higher-order singularities have been ruled out as their existence implies  
a singular pressure in the neck region.

\subsection{Exterior region (wavelength scale)}\label{sec:exterior off resonance}
Consider next the fluid region exterior to the resonator. Far from the neck opening, the latter shrinks to a point such that the only relevant length scale is the wavelength. This suggests considering the ``outer'' external region associated with the stretched coordinate
\begin{equation}\label{outer stretch}
\mathbf{X}^+=\epsilon^{1/2}\bx^+
\end{equation}
and pressure field 
$Q(\mathbf{X}^+) = p(\bx^+)$, where $\unit\bcdot\mathbf{X}^+>0$ as shown in Fig.~\ref{domains}(b). The latter is governed by
\begin{equation}\label{helmholtz exterior}
\nabla_{\mathbf{X}^+}^2 Q + \Omega^2 Q = 0 \quad \text{for} \quad \unit\bcdot\mathbf{X}^+>0,
\end{equation}
where the Laplacian is with respect to the position vector $\mathbf{X}^+$; the Neumann boundary condition
\begin{equation}
\bn\bcdot\bnabla_{\mathbf{X}^+} Q =0 \quad \text{for} \quad \unit\bcdot\mathbf{X}^+=0\setminus \{\mathbf{X}^+ = \bf{0}\};
\end{equation} 
and the condition that the scattered field $Q(\mathbf{X}^+)-Q\ub{i}(\mathbf{X}^+)$ radiates away from the surface, where $Q\ub{i}(\mathbf{X}^+)=\exp(-i\Omega \unit\bcdot\mathbf{X}^+)$. 

The general solution to the above outer problem follows immediately from \eqref{farfie}:
\begin{equation}\label{pressure exterior off}
Q = 2\cos(\Omega \unit\bcdot \mathbf{X}^+) + \epsilon^{1/2}A\frac{\exp\left(i \Omega |\mathbf{X}^+|\right)}{|\mathbf{X}^+|} + \epsilon \boldsymbol{B}\cdot \bnabla_{\mathbf{X}^+}\frac{\exp\left(i \Omega |\mathbf{X}^+|\right)}{|\mathbf{X}^+|} + \cdots. 
\end{equation}
The behavior of $Q(\mathbf{X}^+)$ as $\mathbf{X}^+\to\bzero$, and hence the scalings and values of the coefficients in this exact representation, is to be determined via matching. A preliminary scaling result can be deduced based on our assumption, to be verified, that the pressure is $O(1)$ in the neck region. Since the neck dimensions are $O(\epsilon^{3/2})$ relative to the wavelength, inspection of the magnitude of the second term in \eqref{pressure exterior off} as $\mathbf{X}^+\to\bzero$ shows that $A=O(\epsilon)$. We accordingly write
\begin{equation}\label{A exp offres}
A=\epsilon A_1 + \cdots \quad \text{as} \quad \epsilon\to0,
\end{equation}
where $A_1$ is independent of $\epsilon$. It follows from the same argument that the coefficients of the remaining singular terms in \eqref{pressure exterior off}, starting with $\boldsymbol{B}$, are all $o(\epsilon)$.

\subsection{Neck region}\label{sec:neck off resonance}
The neck region linking the outer and cavity regions is associated with the stretched coordinate 
\begin{equation}\label{neck coord}
\boldsymbol{\xi}=\bx/\epsilon. 
\end{equation}
The neck pressure is accordingly written as $q(\boldsymbol{\xi}) = p(\bx)$. 
The limiting geometry of the neck region is illustrated in Fig.~\ref{domains}(c). The corresponding fluid domain is denoted by $\mathcal{N}$ and consists of the region external to an infinite wall of thickness $2h$ with a cylindrical perforation of radius unity. The domain $\mathcal{N}$ is independent of $\epsilon$ when defined in terms of $\boldsymbol{\xi}$.

The neck problem consists of 
\begin{equation}
\nabla_{\boldsymbol{\xi}}^2 q + \epsilon^3 \Omega^2 q = 0 \quad \text{in} \quad \mathcal{N};
\end{equation}
the Neumann boundary condition
\begin{equation}
\bn\bcdot\bnabla_{\boldsymbol{\xi}}q=0 \quad \text{on} \quad \partial\mathcal{N};
\end{equation}
and asymptotic matching with the exterior and neck regions in the limits $|\boldsymbol{\xi}|\to\infty$ with $\unit\bcdot\boldsymbol{\xi}\gtrless 0$, respectively.

We expand the pressure in the neck region as
\begin{equation}
q = q_0 +\cdots \quad \text{as} \quad \epsilon\to0,
\end{equation}
with $\boldsymbol{\xi}$ fixed, 
where $q_0$ satisfies 
\begin{equation}\label{equation off q0}
\nabla_{\boldsymbol{\xi}}^2 q_{0} = 0 \quad \text{in} \quad \mathcal{N}
\end{equation}
and 
\begin{equation}\label{neumann off q0}
\bn\bcdot\bnabla_{\boldsymbol{\xi}}q_0=0 \quad \text{on} \quad \partial\mathcal{N}.
\end{equation}
Straightforward leading-order asymptotic matching with the cavity region gives
\begin{equation}\label{infinity off q0 cavity}
q_{0} \to \bar{p}_0 \quad \text{as} \quad |\boldsymbol{\xi}|\to\infty \quad (\unit\bcdot\boldsymbol{\xi}<0), 
\end{equation}
whereas from leading-order matching with the exterior region we obtain
\begin{equation}\label{infinity off q0 exterior}
q_{0} \to 2 \quad \text{as} \quad |\boldsymbol{\xi}|\to\infty \quad (\unit\bcdot\boldsymbol{\xi}>0). 
\end{equation}

Equations \eqref{equation off q0}--\eqref{infinity off q0 exterior} describe a potential flow through the neck driven by the pressure difference between the cavity and exterior regions. This type of problem will appear multiple times throughout the paper. Accordingly, it is convenient to introduce a canonical neck field $G(\boldsymbol{\xi})$ that satisfies
\begin{equation}\label{equation  G}
\nabla_{\boldsymbol{\xi}}^2 G = 0 \quad \text{in} \quad \mathcal{N};
\end{equation}
the Neumann boundary condition
\begin{equation}\label{neumann G}
\bn\bcdot\bnabla_{\boldsymbol{\xi}}G=0\quad \text{on} \quad \partial\mathcal{N};
\end{equation}
and 
\begin{equation}\label{conditions G}
G\to \pm\frac{1}{2} \quad \text{as} \quad |\boldsymbol{\xi}| \to\infty \quad (\unit\bcdot\boldsymbol{\xi}\gtrless 0).
\end{equation}
For matching purposes, we are mainly interested in the expansion of $G$ for large $|\boldsymbol{\xi}|$ , which can be written as 
\begin{equation}\label{G beta}
G\sim \pm\frac{1}{2} \mp \frac{\beta}{2\pi |\boldsymbol{\xi}|} + O\left(|\boldsymbol{\xi}|^{-2}\right) \quad \text{as} \quad |\boldsymbol{\xi}| \to\infty \quad (\unit\bcdot\boldsymbol{\xi}\gtrless 0).
\end{equation}
From the above definition, it is straightforward to show that $\beta$ is a real positive function of $h$. It is identical to the Rayleigh ``conductivity'' --- in fact the inverse of the acoustic reactance --- of the neck normalized by its radius \cite{rayleigh:1871,tuck:1975}.

\begin{figure}[t]
\begin{center}
\includegraphics[trim={0 0.5cm 0 0},scale=0.7]{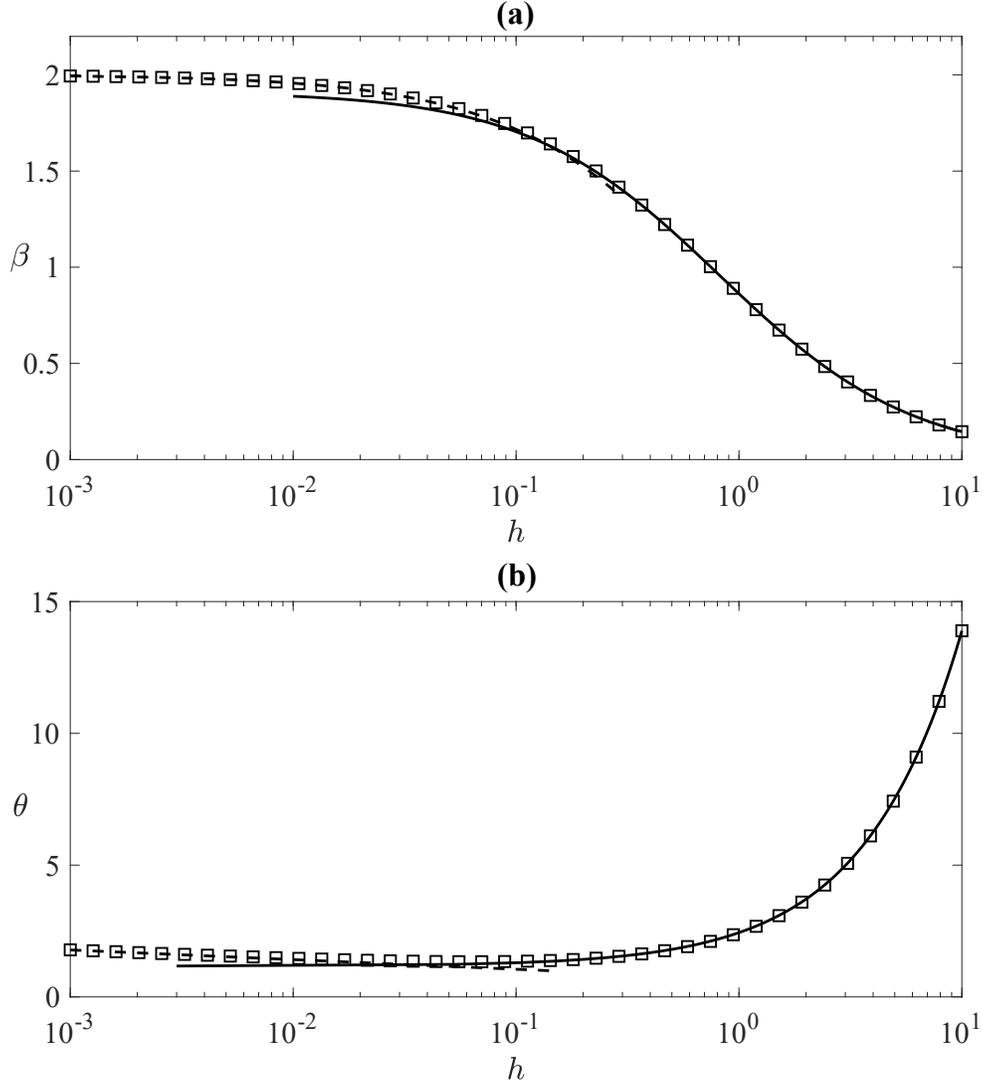}
\caption{The neck parameters (a) $\beta$ and  (b) $\theta$ as a function of $h$. Dashed lines: small-$h$ approximations. Solid lines: large-$h$ approximations. Symbols: numerically calculated parameters. The details of the analysis and numerical methods employed are described in \cite{brandao:20xx}.}\label{fig:betanmu}
\end{center}
\end{figure}
In a recent paper \cite{brandao:20xx}, we derived asymptotic formulas for $\beta$ in the limits of small and large neck aspect ratio $h$:
\begin{equation}\label{beta approx}
\beta \sim  2 - \frac{2}{\pi}h\left(\ln \frac{\pi}{h} + 1\right) + o(h) \quad \text{as} \quad h\to0, \quad \beta \sim \frac{\pi}{2(h + \varpi)} + e.s.t. \quad \text{as} \quad h\to\infty,
\end{equation}
where the numerical constant $\varpi = 0.8217$ can be interpreted as an effective end correction and e.s.t.~stands for exponentially small terms, i.e., terms that are beyond all orders in $h$. In Fig.~\ref{fig:betanmu}(a), we plot the numerically calculated variation of $\beta$ with $h$ alongside the asymptotic expressions for large and small $h$. The symbols are based on a
numerical solution of \eqref{equation  G}--\eqref{G beta} as discussed in \cite{brandao:20xx}.

For the purpose of higher-order matching in later stages of the analysis, it is convenient to rewrite expansion \eqref{G beta} in terms of the shifted position vectors $\boldsymbol{\xi}^{\pm}= \boldsymbol{\xi}\mp h\unit$. This gives
\begin{equation}\label{g no dipole}
G\sim \pm\frac{1}{2} \mp \frac{\beta}{2\pi |\boldsymbol{\xi}^{\pm}|} + O\left(|\boldsymbol{\xi}^{\pm}|^{-3}\right) \quad \text{as} \quad |\boldsymbol{\xi}^{\pm}| \to\infty \quad (\unit\bcdot\boldsymbol{\xi}^{\pm}\gtrless 0),
\end{equation}
where axial symmetry and \eqref{neumann G} were used to eliminate the possibility of $O(|\boldsymbol{\xi}^{\pm}|^{-2})$ dipole terms. In turn, \eqref{g no dipole} can be used to deduce the non-vanishing dipole terms in \eqref{G beta}.

Returning to the problem in \eqref{equation off q0}--\eqref{infinity off q0 exterior}, it is straightforward to solve for $q_0$ in terms of the canonical field $G$:
\begin{equation}
q_0 = (2-\bar{p}_0)G+\frac{2+\bar{p}_0}{2}. 
\end{equation}

\subsection{Leading-order approximation}\label{results off-resonance}
We carry out asymptotic matching using Van Dyke's matching rule \cite{Van:pert}. In the context of the present analysis, it is important to note that this rule should be applied with the inner-outer coordinates related by a pure dilation, rather than both a dilation and a small translation. Accordingly, we match the neck and exterior expansions written in terms of the pair of position vectors $\{\boldsymbol{\xi}^{+},\mathbf{X}^{+}\}$. Similarly, we match the neck and cavity expansions written in terms of the pair of position vectors $\{\boldsymbol{\xi}^{-},\bx^{-}\}$. In this way, and using \eqref{g no dipole}, we obtain $\bar{p}_{0}$ by matching the neck expansion to $O(1)$ with the cavity expansion to $O(\epsilon)$, and $A_1$ by matching the neck expansion to $O(1)$ with the exterior field to $O(\epsilon^{3/2})$. Thus,
\begin{equation}\label{off-resonance results}
\bar{p}_0 = -\frac{2\beta}{\Omega^2 - \beta}, \qquad A_{1} = -\frac{1}{\pi}\frac{\beta\Omega^2}{\Omega^2 -\beta}.
\end{equation}

Since $\beta>0$, the cavity pressure $\bar{p}_0$ and diffraction amplitude $A_1$ are both singular at the resonance frequency
\begin{equation}\label{leading resonance}
\Omega=\beta^{1/2}.
\end{equation}
Thus, the analysis in this section applies only to frequencies sufficiently far from resonance. In the following two sections, we specify the range of validity of the  present ``off-resonance'' analysis precisely and obtain new approximations for the cavity pressure and diffraction amplitude which are valid for frequencies near resonance.

\section{Near-resonance limit: Cavity-shape effect}\label{sec:near resonance}

In the absence of dissipation, the acoustical response of a Helmholtz resonator is limited solely by radiation damping. Thus, the blow-up predicted by the off-resonance analysis of \S\ref{sec:leading} is a consequence of the negligible role of radiation damping in that limit process. Nevertheless, the rate of divergence of the off-resonance approximations \eqref{off-resonance results} as $\Omega\to\beta^{1/2}$ can be used to estimate the width of the frequency interval about the resonance wherein radiation damping is appreciable. Indeed, it is readily seen from \eqref{pressure exterior off} and \eqref{off-resonance results} that the amplitude of the diffracted spherical wave jumps to leading $O(1)$ in the exterior limit when $\Omega-\beta^{1/2}=O(\epsilon^{3/2})$.

It turns out that the off-resonance analysis is actually invalid in a wider, $O(\epsilon)$, frequency interval about the resonance. Indeed, for $\Omega - \beta^{1/2} = O\left(\epsilon\right)$, \eqref{off-resonance results} suggests that the pressure in the cavity jumps to $O\left(\epsilon^{-1}\right)$; as we shall see, in that case the flow inside the cavity becomes important in determining the response of the resonator. In the present section, we focus on this ``near-resonance" $O\left(\epsilon\right)$ frequency interval and leave the ``on-resonance'' analysis of the $O\left(\epsilon^{3/2}\right)$ interval for the next section. 

To study the near-resonance regime we consider a modified limit process where we write $\Omega$ as
\begin{equation}
\label{Ocor}
\Omega = \beta^{1/2} + \epsilon \Omega'
\end{equation}
and hold $\Omega'$ fixed as $\epsilon\to0$.

\subsection{Cavity region}\label{sec:cavity near resonance}
In light of the above, we expand the cavity pressure as
\begin{equation}\label{p near expansion}
p = \epsilon^{-1}{\bar{p}}_{-1} +  p_0 + \epsilon p_{1} +\cdots \quad \text{as} \quad \epsilon\to0,
\end{equation}
wherein $\bar{p}_{-1}$ is a constant value to be determined. The uniformity of the leading-order term can be argued as in \S\ref{sec:cavity off resonance}.

At $O(1)$ we find
\begin{equation}\label{p0 near condition}
\nabla^2 p_0 =- \beta \bar{p}_{-1} \quad \text{in} \quad \mathcal{C},
\end{equation}
subject to
\begin{equation}\label{p0 near neumann}
\bn\bcdot\bnabla p_{0} = 0 \quad \text{on} \quad \partial\mathcal{C}\setminus \{\bx^{-}=\bzero\}.
\end{equation}
Following the derivation in \S\ref{sec:cavity off resonance}, we find that at the degenerate point $p_0$ satisfies the following monopole-singularity condition:
\begin{equation}\label{p0 near singularity}
p_0\sim -\frac{\beta \bar{p}_{-1}}{2\pi |\bx^-|} + O(1) \quad \text{as} \quad |\bx^-|\to0.
\end{equation}
As in \eqref{p1 off singularity}, dipole and higher-order singularities can be ruled out based on the magnitude of the pressure field in the neck region.

The solution of \eqref{p0 near condition}--\eqref{p0 near singularity}, given $\bar{p}_{1}$, is determined up to an additive constant. It is therefore convenient to decompose $p_{0}$ as 
\begin{equation}\label{p0 sol near}
p_0 = \langle p_0 \rangle + \beta \bar{p}_{-1}g ,
\end{equation}
where the first term represents a uniform mean pressure, with 
\begin{equation}
\langle\cdot\rangle = \int_{\mathcal{C}}(\cdot)\,d^3\bx^{-}
\end{equation}
denoting the volume average over the cavity domain (recall that lengths are normalized such that the volume of $\mathcal{C}$ is unity); the second term then accounts for spatial variations of $p_0$ from its mean value. 

By construction, the field $g$ satisfies
\begin{equation}\label{g eq}
\nabla^2 g =-1 \quad \text{in} \quad \mathcal{C};
\end{equation}
the boundary condition
\begin{equation}\label{g neu}
\bn\bcdot\bnabla g = 0  \quad \text{on} \quad \partial\mathcal{C}\setminus \{\bx^{-}=\bzero\};
\end{equation}
the singularity condition at the degenerate point
\begin{equation}\label{g singularity}
g\sim -\frac{1}{2\pi |\bx^-|} + O(1) \quad \text{as} \quad |\bx^-|\to0;
\end{equation}
and the zero-mean condition
\begin{equation}\label{g mean}
\langle g \rangle  = 0,
\end{equation}
which ensures uniqueness.

For matching purposes, we note that \eqref{p0 sol near} and \eqref{g singularity} provide the asymptotic behavior
\begin{equation}\label{p0 near singularity better}
p_0- \langle p_0 \rangle \sim \beta \bar{p}_{-1}\left(-\frac{1}{2\pi |\bx^-|} +  \sigma \right)+ \cdots \quad \text{as} \quad |\bx^-|\to0, 
\end{equation}
where $\sigma$ is a constant defined by the limit
\begin{equation}\label{sigma def}
\sigma =\lim_{|\bx^-|\to0} \left(g+\frac{1}{2\pi |\bx^-|}\right).
\end{equation}

We shall refer to $\sigma$ as the shape factor of the cavity. Similar geometric parameters have been defined by other authors \cite{tuck:1975,bigg:1982,mohring:1999}.
\begin{figure}[t!]
\begin{center}
\includegraphics[trim={3cm 1cm 1cm 2cm}, scale=0.35]{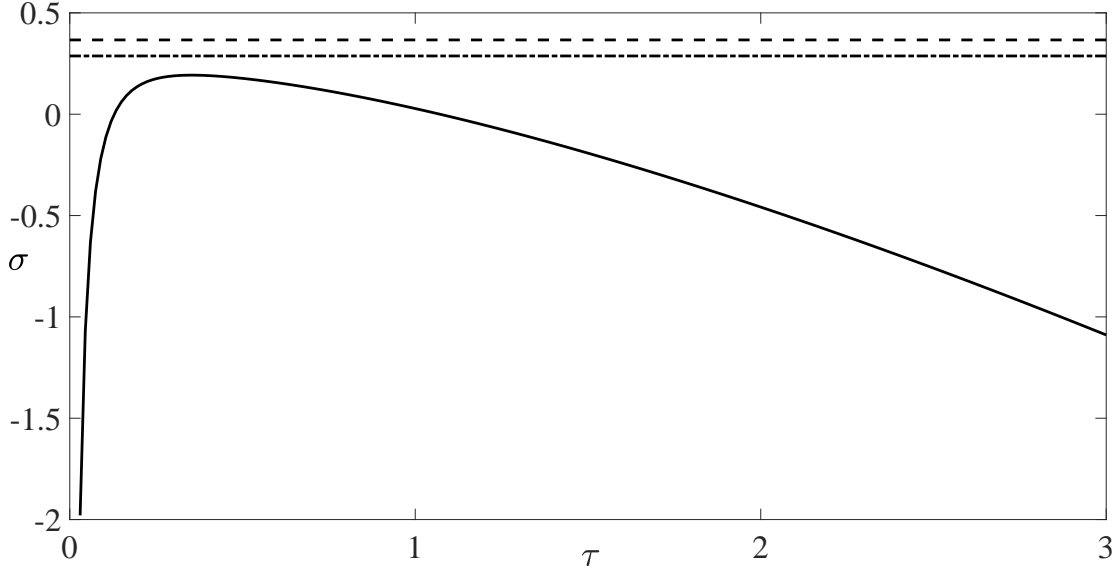}
\caption{Solid line: shape factor $\sigma$ of a cylindrical cavity as a function of its height-to-diameter aspect ratio  $\tau$. Dashed line: $\sigma$ for a hemispherical cavity. Dash-dotted line: $\sigma$ for a cubic cavity.}\label{fig:sigma}
\end{center}
\end{figure}
For a hemispherical cavity, with the singularity \eqref{p0 near singularity better} at the sphere center, a straightforward solution of \eqref{g eq}--\eqref{g mean} yields \cite{bigg:1982,mohring:1999,monkewitz:1985}
\begin{equation}\label{sigma hemisphere}
\sigma_{\text{hemisphere}} = \frac{9}{10\pi}\left(\frac{2\pi}{3}\right)^{1/3}.
\end{equation}
Details are provided in Appendix \ref{sec:csf}, where we also calculate $\sigma$ for a cube and circular cylinder; in both cases, the singularity is located at the center of a face coinciding with the plane $\unit\bcdot\bx^-=0$. For a cube, we find 
\begin{equation}\label{sigma cube}
\sigma_{\text{cube}} \approx 0.2874.
\end{equation}
For a cylinder (aspect ratio $\tau$ between the cylinder's height and diameter),
\begin{equation}\label{sigma cylinder}
\sigma_{\text{cylinder}} = \left(2\pi\tau\right)^{1/3}\left[\frac{1}{4} -  \frac{2\tau}{3 \pi} - \frac{1}{\pi}\sum_{n = 1}^{\infty}\left(  \frac{\coth\left(2\lambda_{n}\tau\right)}{\lambda_{n} J_{0}\left(\lambda_{n} \right)^2} - \frac{\pi}{2}\right)\right].
\end{equation}
In Fig.~\ref{fig:sigma} we plot the variation of $\sigma$ with $\tau$ for a cylinder and compare it to the value for a hemisphere and for a cube. We note that the expression for $\sigma_{\text{cylinder}}$ provided in \cite{bigg:1982} appears to be based on an erroneous solution for the field $g$.

With \eqref{p0 sol near}, the $O(\epsilon)$ cavity problem consists of
\begin{equation}\label{equation near p1}
\nabla^2 p_{1} =- \beta p_{0} - 2\beta^{1/2}\Omega' \bar{p}_{-1} \quad \text{in} \quad \mathcal{C};
\end{equation}
the Neumann condition 
\begin{equation}\label{p1 near neumann}
\bn\bcdot\bnabla p_{1} = 0 \quad \text{on} \quad \partial\mathcal{C}\setminus \{\bx^{-}=\bzero\}\;
\end{equation}
and the singularity condition
\begin{equation}\label{p1 near singularity}
p_1\sim-\frac{\beta \langle p_{0}\rangle +2\beta^{1/2}\Omega'\bar{p}_{-1}}{2\pi |\bx^-|} + O(1) \quad \text{as} \quad |\bx^-|\to0.
\end{equation}
We note that, unlike in preceding asymptotic orders, the absence of an $O(|\bx^-|^{-2})$ dipole-type singularity in \eqref{p1 near singularity} cannot be ruled out based on the magnitude of the pressure in the neck region, which is $O(\epsilon^{-1})$ in the near-resonance regime (though this argument can still be used to rule out higher-order singularities). Nevertheless, matching with the neck region will verify the absence of such a dipolar term.

\subsection{Exterior region (wavelength scale)} 
The $O(\epsilon^{-1})$ pressure enhancement in the cavity suggests a comparable enhancement of the diffraction amplitude $A$, relative to its magnitude off-resonance [cf.~\eqref{pressure exterior off} and \eqref{A exp offres}]. Thus, the diffraction amplitude is expanded as
\begin{equation}
A = A_0 + \cdots \quad \text{as} \quad \epsilon\to0,
\end{equation} 
the corresponding expansion for the exterior pressure being
\begin{equation}\label{pressure near exterior}
Q = 2\cos(\beta^{1/2}\unit\bcdot \mathbf{X}^+) + \epsilon^{1/2}A_{0}\frac{\exp\left(i \beta^{1/2} |\mathbf{X}^+|\right)}{|\mathbf{X}^+|} + \cdots  \quad \text{as} \quad \epsilon\to0.
\end{equation}
Note that the frequency deviation $\epsilon\Omega'$ from resonance does not affect the first two terms of this asymptotic expansion.

\subsection{Neck region}

The pressure in the neck region is also enhanced to $O\left(\epsilon^{-1}\right)$:
\begin{equation}
q = \epsilon^{-1} q_{-1} +   q_{0} +\cdots \quad \text{as} \quad \epsilon\to0.
\end{equation}
Both $q_{-1}$ and $q_{0}$ obey Laplace's equation in $\mathcal{N}$, the usual Neumann condition on $\partial\mathcal{N}$, as well as far-field conditions derived below by matching with the cavity and exterior regions.

Consider first the $O\left(\epsilon^{-1}\right)$ problem governing $q_{-1}$. Straightforward leading-order matching with the pressure in the cavity region gives 
\begin{equation}\label{qm1 far 1}
q_{-1} \to \bar{p}_{-1} \quad \text{as} \quad |\boldsymbol{\xi}|\to\infty \quad (\unit\bcdot\boldsymbol{\xi}<0).
\end{equation}
Similarly, matching with the $O(1)$ pressure in the exterior region gives
\begin{equation}\label{qm1 far 2}
q_{-1} \to 0 \quad \text{as} \quad |\boldsymbol{\xi}|\to\infty \quad (\unit\bcdot\boldsymbol{\xi}>0).
\end{equation}
Thus, $q_{-1}$ is readily expressed in terms of the canonical field $G$ defined in \S\ref{sec:neck off resonance}:
\begin{equation}\label{qm1 solution}
q_{-1} = \bar{p}_{-1}\left(\frac{1}{2}-G\right). 
\end{equation}
Matching the neck pressure to $O(\epsilon^{-1})$ with the exterior pressure to $O(\epsilon^{1/2})$, using \eqref{g no dipole} and \eqref{qm1 solution},
we find 
\begin{equation}
A_{0} = \frac{\beta}{2\pi} \bar{p}_{-1}.
\end{equation}

We now consider the $O\left(1\right)$ problem governing $q_0$. Matching the neck and cavity pressures, both taken to $O(1)$, gives
\begin{equation}
q_{0} \to \langle p_0\rangle +\beta \bar{p}_{-1}\sigma \quad \text{as} \quad |\boldsymbol{\xi}|\to\infty \quad (\unit\bcdot\boldsymbol{\xi}<0).
\end{equation}
An analogous $O(1)$ matching with the exterior pressure gives
\begin{equation}
q_{0} \to 2 \quad \text{as} \quad |\boldsymbol{\xi}|\to\infty \quad (\unit\bcdot\boldsymbol{\xi}>0).
\end{equation}
Thus, $q_0$ is readily expressed as
\begin{equation}
q_0 = \left(2-\langle p_0\rangle - \beta \bar{p}_{-1}\sigma\right)G+\frac{2+\langle p_0\rangle +\beta \bar{p}_{-1}\sigma}{2}. 
\end{equation}

\subsection{Leading-order approximation}\label{resul near}
Matching the neck pressure to $O(1)$ with the cavity pressure to $O(\epsilon)$ yields
\begin{equation}\label{results near}
\bar{p}_{-1} = -\frac{2\beta^{1/2}}{ 2\Omega' - \beta^{3/2}\sigma}, \qquad A_{0} =-\frac{1}{\pi}\frac{\beta^{3/2}}{2\Omega'-\beta^{3/2}\sigma}.
\end{equation}
Matching at these orders also confirms the absence of a dipole-type singularity in \eqref{p1 near singularity}, in turn owing to the absence of such a singular term in \eqref{g no dipole}.
Since $\sigma$ is real, both $\bar{p}_{-1}$ and $A_{0}$ are singular at
\begin{equation}
\label{corrected resonance frequency}
\Omega' = \frac{\beta^{3/2}\sigma}{2},
\end{equation}
corresponding to an $O(\epsilon)$ correction to the leading-order resonance frequency \eqref{leading resonance}.

\section{On-resonance limit: Radiation damping}\label{sec:rdamp}
As radiation damping remains negligible in the near-resonance regime, the singularity of the leading-order approximations \eqref{results near} for the pressure cavity and diffraction amplitude in that limit process is unsurprising. Following the discussion in \S\ref{sec:near resonance}, we next consider the on-resonance regime, which corresponds to an $O(\epsilon^{3/2})$ frequency interval about the corrected resonance frequency $\beta^{1/2}(1+\epsilon\beta\sigma/2)$ 
implied by the near-resonance analysis.

To investigate the on-resonance limit, wherein we expect radiation damping to be important, we rewrite $\Omega$ as 
\begin{equation}\label{omega2 def}
\Omega =  \beta^{1/2}\left(1 + \epsilon\frac{\beta\sigma}{2}\right)+ \epsilon^{3/2} \Omega''
\end{equation}
and consider the limit process $\epsilon\to0$ with $\Omega''$ fixed.

\subsection{Cavity region}\label{sec:cavity on resonance}
In the on-resonance regime, the expansion for the pressure in the cavity becomes
\begin{equation}\label{on expansion}
p = \epsilon^{-3/2} \bar{p}_{-3/2} + \epsilon^{-1}\bar{p}_{-1} + \epsilon^{-1/2}p_{-1/2}+ p_0 + \epsilon^{1/2}p_{1/2} + \epsilon p_{1} +\cdots \quad \text{as} \quad \epsilon\to0,
\end{equation}
wherein $\bar{p}_{-3/2}$ and $\bar{p}_{-1}$ are constants. The argument for the uniformity of the two leading terms is a minor generalization of the one used in \S\ref{sec:cavity off resonance}.

The higher-order terms $p_{-1/2}$ and $p_0$ are both governed by the forced Laplace's equations
\begin{equation}
\nabla^2 p_{-1/2} = - \beta  \bar{p}_{-3/2}, \qquad \nabla^2 p_{0} = - \beta  \bar{p}_{-1} \quad \text{in} \quad \mathcal{C};
\end{equation}
the Neumann conditions
\begin{equation}\label{on resonance neumann}
\bn\bcdot\bnabla p_{{-1/2}} = 0, \quad \bn\bcdot\bnabla p_{{0}} = 0 \quad \text{on} \quad \partial\mathcal{C}\setminus \{\bx^{-}=\bzero\};
\end{equation}
and the singularity conditions
\begin{equation}\label{on resonance singularities}
p_{-1/2}\sim -\frac{\beta\bar{p}_{-3/2}}{2\pi |\bx^-|} + O(1), \qquad p_{0}\sim -\frac{\beta\bar{p}_{-1}}{2\pi |\bx^-|} + O(1) \quad \text{as} \quad |\bx^-|\to0,
\end{equation} 
where dipole and higher-order singularities have been ruled out based on the magnitude of the pressure field in the neck region.
The solutions to both of the above problems is readily expressed in terms of the canonical cavity field $g$:
\begin{equation}\label{on resonance cavity 1}
p_{-1/2}=\langle p_{-1/2}\rangle+\beta \bar{p}_{-3/2}g, \qquad p_{0}=\langle p_0\rangle+\beta \bar{p}_{-1}g. 
\end{equation}

With \eqref{on resonance cavity 1}, we can proceed to consider the higher-order terms $p_{1/2}$ and $p_{1}$. They satisfy the forced Laplace's equations
\begin{equation}
\nabla^2 p_{1/2} = - \beta  p_{-1/2} - \beta^{2} \sigma \bar{p}_{-3/2}, \quad \nabla^2 p_{1} = - \beta  p_0 - \beta^{2} \sigma \bar{p}_{-1} - 2\beta^{1/2}\Omega'' \bar{p}_{-3/2}  \quad \text{in} \quad \mathcal{C};
\end{equation}
the Neumann conditions
\begin{equation}\label{on resonance neumann again}
\bn\bcdot\bnabla p_{{1/2}} = 0, \quad \bn\bcdot\bnabla p_{{1}} = 0 \quad \text{on} \quad \partial\mathcal{C}\setminus \{\bx^{-}=\bzero\};
\end{equation}
and the singularity conditions
\begin{equation}\label{on resonance sing 3}
p_{1/2}\sim -\frac{\beta \langle p_{-1/2}\rangle +\beta^{2}\sigma \bar{p}_{-3/2}}{2\pi |\bx^-|} + O(1)
\end{equation}
and
\begin{equation}\label{on resonance sing 4}
p_1\sim  -\frac{\beta \langle p_{0}\rangle +\beta^{2}\sigma \bar{p}_{-1}+ 2\beta^{1/2}\Omega'' \bar{p}_{-3/2}}{2\pi |\bx^-|} + O(1) \quad \text{as} \quad |\bx^-|\to0.
\end{equation}
The absence of dipole-type singularities in \eqref{on resonance sing 3} and \eqref{on resonance sing 4} will be confirmed by matching. Higher-order singularities have been ruled out based on the magnitude of the pressure field in the neck region.

\subsection{Exterior region (wavelength scale)}\label{sec:exterior on resonance}
Consider next the exterior region. In the on-resonance regime, the diffraction amplitude $A$ becomes $O(\epsilon^{-1/2})$, i.e., 
\begin{equation}
A= \epsilon^{-1/2}A_{-1/2}+\cdots \quad \text{as} \quad \epsilon \to0.
\end{equation}
Thus, in the corresponding expansion for the exterior pressure,
\begin{equation}\label{pressure on exterior}
Q = 2\cos(\beta^{1/2}\unit\bcdot \mathbf{X}^+) + A_{-1/2}\frac{\exp\left(i \beta^{1/2} |\mathbf{X}^+|\right)}{|\mathbf{X}^+|} + \cdots \quad \text{as} \quad \epsilon\to0,
\end{equation}
 the spherical wave is no longer negligible relative to the incident and reflected waves.

\subsection{Neck region}\label{sec:neck on resonance}
The pressure in the neck is expanded as
\begin{equation}\label{expan q on}
q = \epsilon^{-3/2} q_{-3/2} + \epsilon^{-1}q_{-1} + \epsilon^{-1/2}q_{-1/2}+ q_0 +\cdots \quad \text{as} \quad \epsilon\to0,
\end{equation}
in accordance with expansion \eqref{on expansion} for the pressure in the cavity. Each of the first four terms in \eqref{expan q on} satisfies Laplace's equation in $\mathcal{N}$ and the usual Neumann condition on $\partial \mathcal{N}$. The far-field conditions satisfied by these fields are determined below by matching \eqref{expan q on} with expansions \eqref{on expansion} and \eqref{pressure on exterior} for the pressure in the cavity and exterior regions, respectively. 

We first consider the terms $q_{-3/2}$ and $q_{-1}$. Matching the neck and cavity expansions, both taken to $O(\epsilon^{-1})$, yields
\begin{equation}\label{q32 q1 on far 1}
q_{-3/2} \to \bar{p}_{-3/2}, \qquad q_{-1} \to \bar{p}_{-1} \quad  \text{as} \quad |\boldsymbol{\xi}|\to\infty \quad (\unit\bcdot\boldsymbol{\xi}<0),
\end{equation}
whereas the $O(1)$ magnitude of the exterior pressure implies
\begin{equation}\label{q32 q1 on far 2}
q_{-3/2} \to 0, \qquad q_{-1} \to 0 \quad \text{as} \quad |\boldsymbol{\xi}|\to\infty \quad (\unit\bcdot\boldsymbol{\xi}>0).
\end{equation}
Given the far-field conditions \eqref{q32 q1 on far 1} and \eqref{q32 q1 on far 2}, we can express $q_{-3/2}$ and $q_{-1}$ in terms of the canonical neck field $G$: 
\begin{equation}\label{q32 q1 on}
q_{-3/2} = \bar{p}_{-3/2}\left(\frac{1}{2}-G\right), \qquad q_{-1} = \bar{p}_{-1}\left(\frac{1}{2}-G\right).
\end{equation}
Furthermore, matching the neck pressure to $O(\epsilon^{-3/2})$ and the exterior pressure to $O(1)$, using the first of \eqref{q32 q1 on},  gives
\begin{equation}\label{A0}
A_{-1/2}=\frac{\beta}{2\pi} \bar{p}_{-3/2}.
\end{equation}

Consider next the higher-order terms $q_{-1/2}$ and $q_0$. Matching the neck and cavity expansions, both taken to $O(1)$, yields
\begin{equation}\label{q12 on far 1}
q_{-1/2} \to\langle p_{-1/2}\rangle +\beta \bar{p}_{-3/2}\sigma, \quad q_{0} \to\langle p_{0}\rangle +\beta \bar{p}_{-1}\sigma \quad \text{as} \quad |\boldsymbol{\xi}|\to\infty \quad (\unit\bcdot\boldsymbol{\xi}<0),
\end{equation}
whereas matching the neck and exterior expansions, both taken to $O(1)$, yields
\begin{equation}\label{q12 on far 2}
q_{-1/2} \to 0, \quad 
q_{0} \to2 + i\frac{\beta^{3/2}}{2\pi}\bar{p}_{-3/2} \quad \text{as} \quad |\boldsymbol{\xi}|\to\infty \quad (\unit\bcdot\boldsymbol{\xi}>0),
\end{equation}
where we used \eqref{A0}. Given the far-field conditions \eqref{q12 on far 1} and \eqref{q12 on far 2}, we can express $q_{-1/2}$ and $q_0$ in terms of the canonical neck field $G$:
\begin{equation}\label{q12 on}
q_{-1/2} = \left(\langle p_{-1/2}\rangle + \beta \sigma \bar{p}_{-3/2}\right)\left(\frac{1}{2}-G\right),
\end{equation}
\begin{equation}
q_0 = \left(2 + i\frac{\beta^{3/2}}{2\pi} \bar{p}_{-3/2}-\langle p_0\rangle - \beta \sigma  \bar{p}_{-1}\right)G+\frac{2+ i\frac{\beta^{3/2}}{2\pi} \bar{p}_{-3/2}+\langle p_0\rangle +\beta \sigma \bar{p}_{-1} }{2}. 
\end{equation}

\subsection{Leading-order approximation}\label{sec:lo resonance}
It is now possible to match the neck expansion to $O(1)$ and the cavity expansion to $O(\epsilon)$, to find $\bar{p}_{-3/2}$ and $A_{-1/2}$ [cf.~\eqref{A0}]:
\begin{equation}\label{on results}
\bar{p}_{-3/2} = -\frac{2\beta^{1/2}}{2\Omega'' + i\frac{\beta^{2}}{2\pi}}, \qquad A_{-1/2}= -\frac{1}{\pi}\frac{\beta^{3/2}}{2\Omega'' + i\frac{\beta^{2}}{2\pi}}.
\end{equation}
In turn, the same matching procedure also confirms the absence of dipole-type singularities in \eqref{on resonance sing 3} and \eqref{on resonance sing 4}.

The imaginary term in the denominators of \eqref{on results}, which represents radiation damping, ensures that the cavity pressure and diffraction amplitude remain bounded at resonance. Accordingly, it is unnecessary to consider narrower frequency intervals about the resonance. Note that $\bar{p}_{-3/2}$ and $A_0$, given by \eqref{on results}, attain their maximum magnitude at $\Omega''=0$. 

\section{Dissipative thermoviscous effects}\label{sec:dissipation}

\subsection{Thermoviscous model}\label{tv model}
Until now we have been analyzing the lossless model formulated in \S\ref{sec:pf}. In the absence of losses, the sole damping mechanism is energy leakage via radiation. The analysis in \S\ref{sec:rdamp} reveals that this mechanism is weak, resulting in a sharp resonance --- the cavity pressure is $O(\epsilon^{-3/2})$ in a frequency interval of width $O(\epsilon^{3/2})$. It is often the case, however, that dissipation, rather than radiation damping, is dominant. Or it may be the case that both damping mechanisms play a comparable role.
Accordingly, the goal of this section is to extend the lossless results of the preceding sections by taking into account thermoviscous dissipation effects. As we shall see, it will be possible to build on the lossless analysis, as well as our recent analysis  of the acoustic impedance of a cylindrical neck \cite{brandao:20xx}. 

We continue to consider the scenario described in \S\ref{sec:pf}, where an isolated Helmholtz resonator embedded in a rigid substrate is subjected to a normally incident plane wave. We also adopt the same notational conventions as in that section. We assume that the fluid is an ideal gas (equilibrium density $\rho$, specific heat capacities $c_p$ and $c_v$, kinematic viscosity $\nu$ and heat conductivity $\kappa$), in which case losses are associated with viscous and thermal dissipation. The extended model involves three dimensionless time-harmonic fields: the pressure $p$, the velocity field $\bv$ and temperature deviation $T$ (respectively normalized by $p_{\infty}$, $p_{\infty}/\epsilon^{1/2} c \rho$ and $p_{\infty}/c_{p}\rho$). 

The fields $p$, $\bv$ and $T$ satisfy the momentum equation
\begin{equation}
\label{ta1}
-i\Omega\bv = -\bnabla p + \delta^2\left[\nabla^2 \bv + \frac{1}{3}\bnabla(\bnabla\bcdot\bv)\right],
\end{equation}
the continuity equation
\begin{equation}
\label{ta2}
-i\epsilon \Omega p - i\epsilon\left(\gamma - 1\right)\Omega(p - T) + \bnabla\bcdot\bv = 0
\end{equation}
and the energy equation
\begin{equation}
\label{ta3}
i\Omega \left(p - T \right) = \frac{\delta^2}{\Pr} \nabla^2 T
\end{equation}
(see, e.g.,~\cite{pierce:1989}). Three new dimensionless parameters appear in \eqref{ta1}--\eqref{ta3}. The parameter $\delta$ is defined as the ratio
\begin{equation}\label{paramet}
\delta = \frac{l_{v}}{l},
\end{equation}
where $l_{v}=\sqrt{\nu l/\epsilon^{1/2}c}$ is a characteristic viscous length scale for the subwavelength regime of interest (note that $l_v=\sqrt{\nu/\omega}$, with $\omega$ at $\Omega=1$). The parameters $\gamma=c_p/c_v$ and $\Pr=\nu c_p \rho/\kappa$ are the adiabatic index and Prandtl number of the gas (for air, $\gamma\approx1.4$ and $\Pr\approx 0.71$). 

The above thermoviscous equations are supplemented by boundary conditions. On the rigid boundary, the velocity satisfies the usual no-slip condition
\begin{equation}\label{no slip}
\bv = 0,
\end{equation}
while the temperature field satisfies the isothermal condition
\begin{equation}\label{no temp}
T = 0.
\end{equation}
The latter condition is based on the assumption that the thermal conductivity of the solid substrate is much larger than that of the gas phase. 

Furthermore, we consider a normally incident plane wave and require the scattered field to radiate away from the substrate. In principle, since plane-wave solutions of \eqref{ta1}--\eqref{ta3} are inhomogeneous, $p_{\infty}$ should be reinterpreted as the pressure of the incident plane wave at the plane $\unit\bcdot\bx^+=0$, say.
Nevertheless, it is evident in the following asymptotic analysis that the slow exponential variation of propagating waves owing to thermoviscous effects is insignificant.

\subsection{Thin boundary layers}\label{ssec:tbl}
Our interest is in the thin-boundary-layer limit $\delta\ll\epsilon\ll1$. In this regime, the momentum and energy equations, respectively \eqref{ta1} and \eqref{ta3}, are singularly perturbed; oscillatory viscous and thermal boundary layers form at the rigid boundary, across which the tangential velocity and temperature rapidly attenuate from their local bulk values so as to satisfy the boundary conditions \eqref{no slip} and \eqref{no temp}. 
These boundary layers, both of characteristic thickness $\delta$, are assumed thin
compared to the smallest geometric features of the resonator, including the $O(\epsilon)$ neck radius and height. 

In Appendix \ref{app:energy} we carry out a detailed scaling analysis of the thermoviscous model and arrive at several conclusions that immensely simplify the asymptotic analysis later in this section. On the basis of energy-dissipation considerations, we show that the condition $\delta\ll \epsilon$ is tantamount to the condition that the resonance remains weakly damped, meaning that the cavity and neck pressure fields become asymptotically large in a narrow frequency interval. Furthermore, we show that whenever this condition is met, loss is necessarily dominated by viscous dissipation in the boundary layer formed in the subwavelength neck domain. In particular, there are three cases to consider: (i) $\delta\ll \epsilon^{5/2}$: the resonance is radiation limited, the cavity pressure at resonance being $O(\epsilon^{-3/2})$. (ii) $\delta=O(\epsilon^{5/2})$: this represents a distinguished limit where viscous dissipation and radiation damping both play a significant role in limiting the resonance and the rate of energy dissipation is maximized; the cavity pressure remains $O(\epsilon^{-3/2})$. (iii) $\epsilon^{5/2}\ll\delta\ll\epsilon$: the resonance is dissipation limited; the scaling of the cavity pressure at resonance diminishes to $O(\epsilon/\delta)$. 

To analyze the thin-boundary-layer limit, we shall adopt the matched asymptotics formalism of the lossless analysis, with the cavity, exterior and neck regions defined as before. In addition, however, there is an $O(\delta)$ boundary layer adjacent to the rigid boundary; the cavity, exterior and neck regions are accordingly reinterpreted as bulk regions. The exact boundary conditions \eqref{no slip} and \eqref{no temp} are satisfied by the boundary-layer fields, while the bulk fields satisfy effective boundary conditions derived by matching with the boundary layer.

Further scaling arguments given in Appendix \ref{app:energy}, based on direct inspection of \eqref{ta1}--\eqref{no temp}, allow pinpointing the asymptotic order at which the viscous and thermal boundary layers modify the pressure field in the bulk neck, cavity and exterior regions. (Bulk viscothermal effects are subdominant to those boundary-layer effects.) This provides an alternative perspective to the scalings derived based on energy-dissipation principles, which is useful for justifying the matched asymptotics analysis. In particular, it is shown that the viscous boundary layer modifies the neck, cavity and exterior pressure fields at orders $\epsilon^{-1}\delta P$, $\epsilon\delta P$ and $\epsilon^{1/2}\delta$, respectively, where $P$ is the scale of the pressure in the bulk cavity and neck regions (note that $P$ depends on $\delta$ and $\Omega$). Similarly, the thermal boundary layer modifies the neck, cavity and pressure fields at orders $\epsilon^2\delta P$, $\epsilon \delta P$ and $\epsilon^{1/2}\delta$.

\subsection{Radiation-limited resonance}\label{ssec:radlimited}
In the regime $\delta\ll \epsilon^{5/2}$, the resonance is radiation limited and the order of the cavity pressure at resonance is $P=\epsilon^{-3/2}$. In fact, using the above scalings it is readily verified that the leading-order results of the lossless analysis hold. In particular, in the on-resonance frequency interval, the boundary layers generate pressure modifications in the neck, cavity and exterior regions at orders $\epsilon^{-5/2}\delta$, $\epsilon^{-1/2}\delta$ and $\epsilon^{1/2}\delta$, respectively. In comparison, the lossless on-resonance analysis of \S\ref{sec:rdamp} involved the pressure field only up to orders $1,\epsilon$ and $1$, in those respective regions. 
%In the next subsection, we extend the lossless analysis to the borderline regime (ii), where dissipation and radiation damping are both important. As briefly discussed in subsection \S\S\ref{ssec:stronger}, the results of that analysis effectively contain those of regimes (i) and (iii).

\subsection{Distinguished limit: Resonance limited by both loss and radiation damping}
\label{ssec:dist}
We next consider the distinguished limit $\delta=O(\epsilon^{5/2})$, where radiation damping and  dissipation in the neck region play a comparable role in limiting the resonance. It suffices to consider this distinguished limit in the on-resonance frequency interval $\Omega-\beta^{1/2}=O(\epsilon^{3/2})$, wherein $P=\epsilon^{-3/2}$, as it is easily verified that the results of the lossless analysis remain unchanged in the off- and near-resonance intervals. To this end, we define the rescaled viscous parameter 
\begin{equation}\label{thickness bl}
\tilde{\delta} = \epsilon^{-5/2} \delta
\end{equation}
and in what follows consider the limit $\epsilon\to0$ with both $\tilde{\delta}$ and $\Omega''$  [cf.~\eqref{omega2 def}] fixed.

For $P=\epsilon^{-3/2}$ and $\delta=O(\epsilon^{5/2})$, the scalings given in \S\S\ref{ssec:tbl} imply  that the viscous boundary layer in the neck region modifies the bulk pressure at order unity, thus invalidating the lossless on-resonance analysis. In contrast, thermal effects in the neck modify the pressure only at $O(\epsilon^3)$, whereas viscous and thermal effects in the cavity modify the pressure at $O(\epsilon^2)$, which is $O(\epsilon)$ smaller than the highest order included in the lossless on-resonance cavity expansion. Accordingly, the analyses of the cavity and exterior regions in  \S\ref{sec:cavity on resonance} and \S\ref{sec:exterior on resonance}, respectively, remain valid.

In the neck region, the bulk pressure field is expanded as in \eqref{expan q on}:
\begin{equation}\label{expan  neck visc}
q = \epsilon^{-3/2} q_{-3/2} + \epsilon^{-1}q_{-1} + \epsilon^{-1/2}q_{-1/2}+ q_0 +\cdots \quad \text{as} \quad \epsilon\to0.
\end{equation}
Writing \eqref{ta1}--\eqref{ta3} in terms of the stretched neck coordinates \eqref{neck coord} and rescaled viscous parameter \eqref{thickness bl}, it is readily verified that all the terms included in \eqref{expan  neck visc} satisfy Laplace's equation in $\mathcal{N}$, as they do in the lossless case. Furthermore, since the solution in the cavity and exterior regions retain the same form as in the lossless case, the matching conditions \eqref{q32 q1 on far 1}--\eqref{q32 q1 on far 2} and \eqref{q12 on far 1}--\eqref{q12 on far 2}  at $|\boldsymbol{\xi}|\to\infty~(\unit\bcdot\boldsymbol{\xi}\gtrless 0)$ remain valid. The difference from the lossless analysis lies in the boundary conditions on $\partial\mathcal{N}$, which must be derived by matching the bulk neck region with a boundary-layer region of thickness $O(\epsilon^{5/2})$ [$O(\epsilon^{3/2})$ relative to the neck scale]. A similar procedure was carried out in the appendix of our recent paper on the acoustical impedance of cylindrical necks \cite{brandao:20xx}. Translating the results in that appendix into the present context, we find that $q_{-3/2}$, $q_{-1}$ and $q_{-1/2}$ satisfy the usual homogeneous Neumann boundary condition on $\partial\mathcal{N}$, whereas $q_0$ satisfies 
the inhomogeneous boundary condition
\begin{equation}\label{neumann visc}
\bn\bcdot\bnabla_{\boldsymbol{\xi}}q_{0}= -\frac{1 + i}{2^{1/2} \beta^{1/4}} \tilde{\delta}\bar{p}_{-3/2}\nabla^2_{s} G \quad \text{on} \quad \partial\mathcal{N},
\end{equation}
where $\nabla^2_{s}$ is the surface Laplacian operator with respect to the position vector $\boldsymbol{\xi}$ \cite{van:2007}. The right-hand side of \eqref{neumann visc} represents the displacement by the boundary layer of the leading-order inviscid flow.

It follows that the fields $q_{-3/2}$, $q_{-1}$ and $q_{-1/2}$ can be expressed in terms of the corresponding cavity pressure fields and the canonical neck field as in \eqref{q32 q1 on} and \eqref{q12 on}. Furthermore, matching between the neck and the cavity gives the diffraction amplitude $A_{-1/2}$, defined by \eqref{pressure on exterior}, as in \eqref{A0}. As anticipated, only the field $q_0$ needs to be revisited.

Fortunately, it is not necessary to explicitly solve the problem governing $q_{0}$, as only the behavior of this field as  $|\boldsymbol{\xi}|\to\infty$ is required for the final matching procedure that furnishes the requisite leading-order approximation for the cavity pressure and diffraction amplitude. To determine this behavior, we shall follow \cite{brandao:20xx} in deriving a reciprocal relation between the field $q_0$ and the canonical neck field $G$ [cf.~\S\S\ref{sec:neck off resonance}].

To this end, we use \eqref{q12 on far 1} and \eqref{q12 on far 2} to write the far fields of $q_0$ in the form
\begin{align}\label{visc far 1}
q_{0} &\sim  \langle p_{0}\rangle +\beta \bar{p}_{-1}\sigma + \frac{F}{2\pi|\boldsymbol{\xi}|} +  O\left(|\boldsymbol{\xi}|^{-2}\right) \quad \text{as} \quad |\boldsymbol{\xi}| \to\infty \quad (\unit\bcdot\boldsymbol{\xi}<0),\\
\label{visc far 2}
q_{0} &\sim 2 + i\frac{\beta^{3/2}}{2\pi} \bar{p}_{-3/2} - \frac{F}{2\pi|\boldsymbol{\xi}|} +  O\left(|\boldsymbol{\xi}|^{-2}\right) \quad \text{as} \quad |\boldsymbol{\xi}| \to\infty \quad (\unit\bcdot\boldsymbol{\xi}>0),
\end{align}
where the monopole strength $F$ is an output of the problem governing $q_0$. To determine $F$, we employ Green's identity between $q_0$ and $G$,
\begin{equation}
\oint\left(q_{0}\bnabla G -G\bnabla q_{0}\right)\bcdot \bn \,dA=0,
\end{equation}
where the integral is over the boundary of  the domain consisting of the portion of $\mathcal{N}$ bounded by the ball $|\boldsymbol{\xi}|<\xi$, $\bn$ being the outward normal to that boundary. By evaluating the integral in the limit $\xi\to\infty$, using \eqref{neumann G}, \eqref{G beta}, \eqref{visc far 1} and \eqref{visc far 2}, as well as integration by parts, we find
\begin{equation}\label{F result}
F =  \beta\left(2 + i\frac{\beta^{3/2}}{2\pi}\bar{p}_{-3/2}-\langle p_{0}\rangle -\beta \sigma \bar{p}_{-1} +  \frac{1 + i}{2^{1/2}} \tilde{\delta}\beta^{3/4}\theta \bar{p}_{-3/2} \right),
 \end{equation}
where we define the parameter
\begin{equation}\label{theta def}
\theta = \beta^{-2}\int_{\partial{\mathcal{N}}} |\bnabla_{s}G|^2 dA,
\end{equation}
$\bnabla_s$ being the surface gradient operator with respect to the position vector $\boldsymbol{\xi}$ \citep{van:2007}. 

Like $\beta$, $\theta$ is a real positive function of $h$ that characterizes the geometry of the neck. In \cite{brandao:20xx}, we evaluated the quadrature in \eqref{theta def} using a numerical solution for the canonical field $G$. We also developed asymptotic formulas in the limits of small and large aspect ratio $h$:
\begin{align}\label{theta approx}
\theta \sim \frac{1}{2}\left(\frac{1}{\pi}\ln\frac{\pi}{h}+1\right) + \cdots \quad \text{as} \quad h\to0, \qquad
\theta\sim \frac{4}{\pi}(h+\varpi_{v}) + e.s.t.\quad \text{as} \quad h\to\infty,
\end{align}
where the numerical constant $\varpi_{v} = 0.91$ can be interpreted as a viscous end correction. See Fig.~\ref{fig:betanmu}(b) for a comparison between the asymptotic formulas and the numerical data. The fact that $\theta$ exhibits a minimum as a function of $h$ agrees with the intuition that viscous resistance is enhanced for very short and very long necks; short necks feature sharp edges and hence higher velocities, whereas long necks have a larger surface area \cite{brandao:20xx}.

A disclaimer is necessary regarding the logarithmic divergence of the parameter $\theta$ as $h\to0$. Whereas we have considered $h$ as a constant parameter, this divergence suggests that our theory breaks down for sufficiently small $h$. Indeed, for $h=O(\delta/\epsilon)$, the neck is so short that the viscous boundary layer becomes comparable to the height of the neck (it remains small compared to the neck radius). As discussed in \cite{brandao:20xx}, the analysis needs to be modified in that case to include an additional viscous inner region, distinguished from the boundary layer, at $O(\delta)$ distances from the sharp edge of the neck. When this additional region is accounted for, it is found that $\theta$ effectively saturates at an $O(\ln \delta)$ value.

We can now proceed as in the lossless on-resonance analysis to match the neck expansion to $O(1)$, using \eqref{visc far 1} and \eqref{F result}, with the cavity expansion to $O(\epsilon)$. We thereby find the leading-order cavity pressure and diffraction amplitude as
\begin{equation} \label{p32 on resonance viscous}
\bar{p}_{-3/2} = -\frac{2\beta^{1/2}}{2\Omega'' +\frac{\tilde{\delta}\beta^{5/4}\theta}{2^{1/2}}+  i\left(\frac{\beta^2}{2\pi}+\frac{\tilde{\delta}\beta^{5/4}\theta}{2^{1/2}}\right)}, \quad A_{-1/2} = -\frac{1}{\pi}\frac{\beta^{3/2}}{2\Omega'' +\frac{\tilde{\delta}\beta^{5/4}\theta}{2^{1/2}}+  i\left(\frac{\beta^2}{2\pi}+\frac{\tilde{\delta}\beta^{5/4}\theta}{2^{1/2}}\right)}.
\end{equation}
These expressions generalize the corresponding lossless results \eqref{on results}. Note that the resonance has shifted from $\Omega''=0$ in the lossless case to 
\begin{equation}\label{frequency on resonance viscous}
\Omega'' = -\frac{\tilde{\delta}\beta^{5/4}\theta}{2^{3/2}}.
\end{equation}

\subsection{Dissipation-limited resonance}\label{ssec:stronger}
In the regime $\epsilon^{5/2}\ll\delta\ll\epsilon$,  the resonance is dissipation limited, though it  remains weakly damped. Furthermore, the order of the cavity pressure at resonance decreases to  $P=\epsilon/\delta$; accordingly, the on-resonance frequency interval expands to $\Omega-\beta^{1/2}=O(\delta/\epsilon)$. It is physically meaningful to split this regime into three sub-regimes. (i) $\epsilon^{5/2}\ll\delta\ll\epsilon^2$: the on-resonance interval remains $\ll\epsilon$ and hence the $O(\epsilon)$ frequency correction due to the inviscid flow in the cavity remains important to leading order. (ii) $\delta=O(\epsilon^2)$: this is a distinguished limit where the on-resonance and near-resonance intervals coincide. (iii) $\epsilon^2\ll \delta\ll\epsilon$: the on-resonance interval is wider than the near-resonance frequency interval and hence the $O(\epsilon)$ frequency correction becomes negligible.

It is tedious but straightforward to verify that in all of the above sub-regimes, approximations to the cavity pressure and diffraction amplitude, to leading order and in the on-resonance frequency interval, can be obtained simply by extrapolating  \eqref{p32 on resonance viscous}. 
In particular, in the distinguished limit $\delta=O(\epsilon^2)$, revisiting the lossless near-resonance analysis of \S\ref{sec:near resonance}, with viscous effects in the neck entering at $O(1)$ as in \S\S\ref{ssec:dist}, we find 
\begin{equation} 
\bar{p}_{-1} = -\frac{2\beta^{1/2}}{2\Omega' -\beta^{3/2}\sigma+\frac{\delta\beta^{5/4}\theta}{\epsilon^{2}2^{1/2}}+  i\frac{\delta\beta^{5/4}\theta}{\epsilon^{2}2^{1/2}}}, \quad A_{0} = -\frac{1}{\pi}\frac{\beta^{3/2}}{2\Omega' -\beta^{3/2}\sigma+\frac{\delta\beta^{5/4}\theta}{\epsilon^{2}2^{1/2}}+  i\frac{\delta\beta^{5/4}\theta}{\epsilon^{2}2^{1/2}}}.
\end{equation}

\section{Single resonator: summary and discussion}\label{sec:discussion}
We used matched asymptotic expansions to analyze the acoustical response of a surface-embedded Helmholtz resonator excited by a normally incident plane wave. A central feature of our analysis is the separate consideration of distinguished frequency intervals converging to the resonance, in addition to the usual treatment of distinguished spatial regions using matched asymptotics. Based on this decomposition, and  scaling arguments, the problem of accounting for thermoviscous effects in the  thin-boundary-layer (i.e., weak damping) regime $\delta\ll\epsilon$ was reduced to that of calculating the perturbative effect of the viscous boundary layer on the inviscid flow in the neck region; this perturbation, in turn, was seen to affect the leading-order response of the resonator whenever $\delta\gtrsim \epsilon^{5/2}$.

In the present section we aim to recapitulate the results derived in the preceding sections in a form which is more convenient for applications. In particular, by combining previous results we provide in \S\S\ref{ssec:freq} a high-order approximation for the resonance frequency, including effects of flow in the cavity and viscous dissipation in the neck; and in \S\S\ref{ssec:uniform} a composite model for the acoustical response of the resonator, valid to leading order as $\epsilon\to0$ for $\Omega=O(1)$ and $\delta\ll\epsilon$.
Furthermore, in \S\S\ref{ssec:dis} we calculate the dissipation rate at resonance and derive a constraint on the parameters of the  resonator so that dissipation is optimal. 
Lastly, in \S\S\ref{ssec:arbinc} we discuss a straightforward extension of the model to the case of an arbitrary incident field, as a step towards the extension presented in \S\ref{sec:mult} to the case of multiple resonators and an arbitrary incident field. 

\subsection{Resonance frequency}\label{ssec:freq}
We define the resonance frequency as the frequency at which the magnitude of the cavity pressure attains a maximum. Our analysis in the preceding sections provides a high-order approximation for the resonance frequency [cf.~\eqref{omega2 def} and \eqref{frequency on resonance viscous}]. Denoting the dimensionless frequency $\Omega$ [cf.~\eqref{Dimensionless frequency}] at resonance by $\bar{\Omega}$, this approximation reads
\begin{equation}\label{comb reso}
\bar{\Omega} \approx \beta^{1/2} + \epsilon\frac{\beta^{3/2}\sigma}{2} -\frac{1}{\epsilon}\frac{\delta\beta^{5/4}\theta}{2^{3/2}},\end{equation}
where $\beta$ and $\theta$ are real positive functions of the neck aspect ratio $h$, and $\sigma$ is a real (positive or negative) function of the cavity geometry.  Analytical expressions for these parameters, which are defined through canonical boundary-value problems and quadratures, are summarized in Table \ref{table:parameters} --- see caption for relevant references and equation numbers in the analysis. 
\begin{table}[]
\setlength{\tabcolsep}{12pt} % Default value: 6pt
\begin{tabular}{|c|c|c|}
\hline
\multicolumn{3}{|c|}{\textbf{Cylindrical-neck parameters} (Fig.~\ref{fig:betanmu})}              \\ \hline
\textbf{}                         & $h\ll 1$   & $ h\gg 1$   \\ \hline
$\beta$         \eqref{beta approx}                 &  $2 - \frac{2}{\pi}h\left(\ln \frac{\pi}{h} + 1\right)$  & $\frac{\pi}{2(h + 0.8217\ldots)}$                     \\ \hline
$\theta$      \eqref{theta approx}                & $\frac{1}{2}\left(\frac{1}{\pi}\ln\frac{\pi}{h}+1\right)$ &  $ \frac{4}{\pi}(h+0.91\ldots)$
                   \\ \hline
\multicolumn{3}{|c|}{\textbf{Cavity shape factor $\sigma$} (Fig.~\ref{fig:sigma})} \\ \hline
Hemisphere  \eqref{sigma hemisphere}                      & \multicolumn{2}{c|}{$\frac{9}{10\pi}\left(\frac{2\pi}{3}\right)^{1/3}=0.3665\ldots$}   \\ \hline
Cube   \eqref{sigma cube}                           & \multicolumn{2}{c|}{0.2874\ldots}   \\ \hline
Cylinder \eqref{sigma cylinder}    & \multicolumn{2}{c|}{$\left(2\pi\tau\right)^{1/3}\left[\frac{1}{4} -  \frac{2\tau}{3 \pi} - \frac{1}{\pi}\sum_{n = 1}^{\infty}\left(  \frac{\coth\left(2\lambda_{n}\tau\right)}{\lambda_{n} J_{0}\left(\lambda_{n} \right)^2} - \frac{\pi}{2}\right)\right] \quad (\lambda_n = n\text{th root of } J_1)$}   \\ \hline
\end{tabular}
\caption{Analytical expressions for the lumped parameters characterizing the neck and cavity geometries. The neck parameters $\beta$ and $\theta$ are positive functions of the neck aspect ratio $h$ that are defined respectively by the canonical  problem \eqref{equation G}--\eqref{G beta} and the quadrature \eqref{theta def}. The approximations given above were derived in \cite{brandao:20xx} and provide accurate values for arbitrary $h$ (see Fig.~\ref{fig:betanmu}). The shape factor $\sigma$ is a real function of the cavity geometry, defined by the canonical problem \eqref{g eq}--\eqref{g mean}. The values for  hemispherical, cubical and cylindrical (aspect ratio $\tau$ between the cylinder's height and diameter) cavities are derived in Appendix \ref{sec:csf}; the result for a hemispherical was previously derived in \cite{bigg:1982}. As demonstrated in Fig.~\ref{fig:sigma}, $\sigma$ can be either positive or negative.} 
\label{table:parameters}
\end{table}

The first, leading-order, term in \eqref{comb reso} corresponds to the classical resonance frequency \eqref{dimresf}  with $H=\pi\epsilon l/\beta$ (assuming $S=\pi\epsilon^2 l^2$); for large $h$ this gives $H\approx 2(h+\varpi)\epsilon l$ hence the interpretation of $\varpi=0.8217\ldots$ as an effective length correction.
The second term is associated with the inviscid flow in the cavity of the resonator. This term is equivalent to frequency corrections found in previous lossless analyses based on layer-potential techniques \cite{mohring:1999} and  matched asymptotic expansions \cite{bigg:1982} (the latter in a more cumbersome form containing spurious contributions). The third term, which we believe has not been derived before,
is associated with the viscous boundary layer in the neck region. We note that our scaling arguments confirm that thermal effects, originally included in our problem formulation, are in fact subdominant. While this point has been argued before in the literature \cite{ingard:1953}, it is often overlooked \cite{howe:1976,jimenez:2017}.

In \eqref{comb reso}, the order of the asymptotic error and the asymptotic ordering of terms depend on the relative smallness of the dissipation parameter $\delta$, defined in \eqref{paramet}. In the radiation-limited regime, $\delta\ll\epsilon^{5/2}$, as well as in the distinguished limit $\delta=O(\epsilon^{5/2})$, where the resonance is limited by both radiation damping and thermoviscous losses, the approximation is correct to $O(\epsilon^{3/2})$, though in the former case the third term becomes negligible. In the dissipation-limited regime, $\epsilon^{5/2}\ll\delta\ll\epsilon$, the approximation is correct to $O(\delta/\epsilon)$, the order of the third term; the second term is then negligible for $\epsilon^2\ll\delta$. 

\subsection{Cavity pressure and diffraction amplitude: unified model}\label{ssec:uniform}
The acoustical response of the resonator is characterized by the pressure in the cavity $\bar{p}$, which is uniform to leading order, and the amplitude of the diffracted spherical wave $A$ defined by \eqref{farfie}. 
Combining previous results, we deduce the following composite approximations:
\begin{equation}\label{uniform approx}
\bar{p} = -\frac{2\beta}{\Omega^2 - \bar{\Omega}^2  + i\left(\epsilon^{3/2}\frac{\beta^{5/2}}{2\pi}+ \frac{1}{\epsilon}\frac{\delta\beta^{7/4}\theta}{2^{1/2}}\right)},  \quad A=  \frac{\epsilon \Omega^2}{2\pi}\bar{p}.
\end{equation}

It can be verified that the approximations \eqref{uniform approx} are valid to leading order as $\epsilon\to0$, for $\Omega=O(1)$ and $\delta\ll\epsilon$.  By valid to leading order we mean that the relative error between the exact and approximate solutions vanishes as $\epsilon\to0$,
which does not imply that the approximations are uniformly valid to any certain order. Strictly speaking, since the order of the cavity pressure, say, is largest in the on-resonance frequency interval (whose width depends on $\delta$), it is in a sense inconsistent to consider the leading approximations outside that interval, which are of higher order. Nevertheless, these composite approximations are practically convenient, especially for the extension presented in \S\ref{sec:mult} to multiple resonators. %given that the width of the on-resonance frequency interval expands as $\delta$ increases; moreover, the leading approximation off-resonance is important in its own right if the external sources operate off-resonance, as well as in some scenarios involving interacting non-identical resonators (see \S\ref{sec:mult}).

\subsection{Critical coupling and optimal dissipation}\label{ssec:dis}
Starting from an exact energy-dissipation relation, in Appendix \ref{app:energy} we derived an approximate relation \eqref{optical theorem} 
between the dissipation rate $D$ of the resonator (normalized by $p_{\infty}^2 l^2/\epsilon^{1/2} c \rho$) and the diffraction amplitude $A$, defined in \eqref{farfie}. In particular, this relation implies the approximate upper bound $D\le\pi/\epsilon^{1/2}\Omega^2$. This bound is attained when the dissipation rate is equal to the rate of radiation damping, a condition sometimes referred to as critical coupling. A similar bound for a resonator in free space was given in \cite{ingard:1953}. 
\begin{figure}[b]
\begin{center}
\includegraphics[scale=0.35]{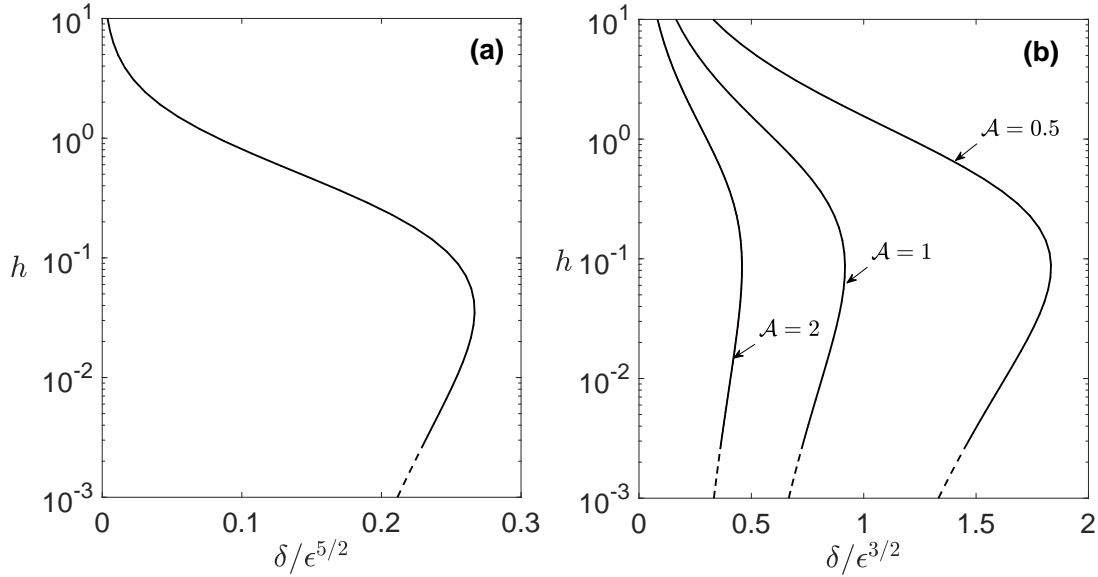}
\caption{Neck aspect ratio $h$ for critical coupling. (a) Single resonator [cf.~\eqref{critc single}]. (b) Subwavelength metasurface for indicated values of the unit-cell area $\mathcal{A}$ [cf.~\eqref{ccoupling}]. Dashed tails hint to breakdown of theory for very short necks, $h=O(\delta/\epsilon)$, as discussed in \S\S\ref{ssec:dis}; the  analysis in \cite{brandao:20xx} suggests  that as $h\to0$ the curves saturate at values of $\delta/\epsilon^{5/2}$ and $\delta/\epsilon^{3/2}$ on the order of $1/\ln\delta$, respectively in the single-resonator and metasurface cases.}
\label{fig:critc}
\end{center}
\end{figure}

Using our asymptotic theory, we can easily determine a condition on the parameters of the resonator such that critical coupling is achieved. Clearly, the upper bound can only be attained in the on-resonance frequency interval. We accordingly calculate the dissipation by substituting our on-resonance approximation \eqref{p32 on resonance viscous} into relation \eqref{optical theorem}. We thereby find that $D$ exhibits a maximum, as a function of frequency, at the resonance frequency \eqref{comb reso}. The dependence of the maximal value upon the resonator parameters is found to be
\begin{equation}
D\approx \frac{4\pi}{\beta\epsilon^{1/2}}\frac{t}{(1+t)^2}, \quad t=\frac{\sqrt{2}\pi{\delta}\theta}{\epsilon^{5/2}\beta^{3/4}},
\end{equation}
Note that this value is independent of the cavity shape factor $\sigma$. Since, $\Omega^2\sim \beta$ at resonance, the above-mentioned upper bound on $D$ is attained for $t=1$. Thus, the critical-coupling condition for a single resonator is
\begin{equation}\label{critc single}
\frac{\delta}{\epsilon^{5/2}}=\frac{\beta^{3/4}}{\sqrt{2}\pi \theta}.
\end{equation}

The right-hand side of \eqref{critc single} is a function of the neck aspect ratio $h$, which is easily calculated using Table \ref{table:parameters}. The left-hand side of \eqref{critc single} is a dimensionless grouping that is independent of $h$; it involves the cavity scale $l$, dimensionless neck size $\epsilon$, kinematic viscosity $\nu$ and the speed of sound $c$ [cf.~\eqref{paramet}]. The critical-coupling condition \eqref{critc single} is depicted in Fig.~\ref{fig:critc}(a).  Note that there is a maximum value of $\tilde{\delta}=\delta/\epsilon^{5/2}$, say $\tilde{\delta}_c$ beyond which critical coupling cannot be achieved for any value of $h$. For $\tilde\delta<{\tilde\delta}_c$, the slow logarithmic divergence of the parameter $\theta$ as $h\to0$ results in a two-valued relation $h(\tilde{\delta})$. As discussed in \S\ref{sec:dissipation}, however, our thermoviscous theory does not hold for neck heights comparable to the viscous boundary layer, $h\lesssim \delta/\epsilon$. In that regime, $\theta$ saturates at $O(\ln \delta)$ values. Thus, in practice, we expect there to exist a second critical value $\tilde{\delta}_a=O(1/\ln \delta)$ such that there are one, two and zero $h$ values satisfying \eqref{critc single} for $0<\tilde{\delta}<\tilde{\delta}_a$, $\tilde{\delta}_a<\tilde{\delta}<\tilde{\delta}_c$ and $\tilde{\delta}>\tilde{\delta}_c$, respectively.

%To concluding: Of course, in practice interest is not attaining maximum dissipation but also broadband, which can be via trade off by increasing $\delta$ as discussed in classical paper by ingard; using explicit systematic the optimised design for any specified goal can easily done without computations or by more sophisticated design for example with multiple resonators of different frequencies. 
%Note that this is $O(\tilde{\delta}/\epsilon^{1/2})$ and this largness is 

\subsection{Scattering factor and extension to arbitrary incident field}\label{ssec:arbinc}
For simplicity, we have so far considered the case of a normally incident plane wave. It is straightforward to extend our results \textit{a posteriori} to virtually arbitrary incident fields; there are special cases where further analysis may be required and this is beyond the scope of the present paper. In particular, we shall assume that the incident field, defined as the field in the absence of the resonator and the rigid wall, is generated by sound sources outside the resonator, at infinity or at least at relatively large distances from the neck opening, i.e., $|\bx^+|\gg\epsilon$ (note that near-field sources positioned at order unity distances from the neck opening are permitted). To be specific, we assume that the incident field is approximately uniform on the scale of the neck region. In that case  the exterior pressure field for $|\bx^+|\gg\epsilon$ can  be approximated as 
\begin{equation}\label{farfie gen}
p = p_{i}(\mathbf{x}^{+}) + p_{r}(\mathbf{x}^{+}) + A\frac{\exp\left(i \epsilon^{1/2} \Omega |\mathbf{x}^{+}|\right)}{|\mathbf{x}^{+}|},
\end{equation} 
where $p_i(\bx^+)$ and $p_{r}(\bx^+)$ denote the incident and reflected fields (thermoviscous losses affect these fields only on scales much larger than the wavelength), and the third term is the spherical wave diffracted from the subwavelength resonator.  Both the incident and reflected fields satisfy the Helmholtz equation \eqref{Helmholtz equation} in the half space $\unit\bcdot\bx^+>0$, except, perhaps, where $p_i(\bx^+)$ is singular; furthermore, the reflected field $p_r(\bx^+)$ is such that the total ambient field $p_i(\bx^+)+p_r(\bx^+)$ satisfies a homogeneous Neumann condition on the plane $\unit\bcdot\bx^+=0$. Note that since the surface is rigid and flat the reflected field can readily be obtained using the method of images \cite{jackson:book}.

It is clear from the analysis in the preceding sections that the response of the resonator is linear in the value of the ambient field at $\bx^+=\bzero$ (the case where the ambient field vanishes there requires special consideration). For example, for the normally incident plane wave considered in \S\ref{sec:pf} that value is $p_i(\bzero)+p_r(\bzero)=2$. It thus follows from  \eqref{uniform approx}, together with linearity, that in the present general scenario the diffraction amplitude $A$ and the cavity pressure $\bar{p}$ are given by
\begin{equation}\label{arb inc ext}
 A = \left(p_{i}(\mathbf{0}) + p_{r}(\mathbf{0})\right) f, \quad \bar{p}=\frac{2\pi}{\epsilon\Omega^2}A,
\end{equation}
where we define the scattering factor
\begin{equation}\label{f}
f = -\frac{\epsilon}{2\pi} \frac{\beta\Omega^2}{\Omega^2 - \bar{\Omega}^2   + i\left(\epsilon^{3/2}\frac{\beta^{5/2}}{2\pi}+ \frac{1}{\epsilon}\frac{\delta\beta^{7/4}\theta}{2^{1/2}}\right)}.
\end{equation}

%\section{Concluding remarks and Generalizations}\label{sec:cr}
%We have developed an asymptotic model of a Helmholtz resonator embedded in a rigid surface. The model is derived from first principles and, as far as we are aware, is unique as it systematically accounts for thermoviscous dissipation. The geometric parameters associated with the model are precisely defined through canonical problems, with explicitly formulas provided for a cylindrical neck and some cavity shapes. Other geometric configurations, such as resonators with non-circular necks, can also be considered by means of trivial generalizations of our model. Our model exhibit a minimal form, which illuminates the role of different physical effects in the response of the resonator, as well as clarifies certain discrepancies present in the literature. We emphasize that our model does not rely on any additional physically-motivated assumption and, as a consequence, all the asymptotic formulas we present directly follow directly from the fundamental equations of acoustics.

\section{Extension to multiple resonators}\label{sec:mult}
\subsection{Foldy's method}\label{ssec:foldymethod}
As a model of an acoustic metasurface, let us consider the case where the rigid substrate is decorated with $N$, not necessarily identical, Helmholtz resonators of the type considered previously. We label the resonators by the index $n=1,2,\ldots,N$. We continue to normalize lengths by $l$,  with $l^3$, $\epsilon l$ and $2\epsilon h l$ now respectively corresponding to the volume, neck radius and neck height of the the first resonator. (Note that $\Omega$ and $\delta$ are defined consistently, \text{viz.}, based on the parameters of the first resonator.) The corresponding quantities of the $n$th resonator are respectively  denoted $v_{(n)} l^3$, $\epsilon a_{(n)} l$ and $2 \epsilon a_{(n)} h_{(n)} l$, with $v_{(1)},a_{(1)}=1$ and $h_{(1)}=h$. We denote the position of the $n$th  resonator by $\bx^{+}=\mathbf{r}_{(n)}$, where we set $\mathbf{r}_{(1)} = \mathbf{0}$ without loss of generality. As in \S\S\ref{ssec:arbinc}, the decorated substrate is externally forced by a general incident field $p_i(\bx^+)$, with associated reflected field $p_r(\bx^+)$ owing to the presence of the rigid substrate.

We assume that $|\br_{(i)}-\br_{(j)}|\gg\epsilon$ for all $i,j=1,2,\ldots,N$. Then at distances $\gg \epsilon$ from the surface, the exterior pressure can be approximated as 
\begin{equation}\label{mult pres}
p = p_{\text{i}}(\mathbf{x}^{+}) + p_{\text{r}}(\mathbf{x}^{+}) + \sum_{n} A_{(n)}\frac{\exp{(i \epsilon^{1/2} \Omega |\mathbf{x}^{+} - \mathbf{r}_{(n)}|)}}{|\mathbf{x}^{+} - \mathbf{r}_{(n)}|},
\end{equation}
where $A_{(n)}$ is the amplitude of the spherical wave diffracted from the $n$th resonator; the pressure inside the cavity of the $n$th resonator is obtained from the diffraction amplitude of that resonator based on the analogous relation for a single resonator:
\begin{equation}\label{w p rel 2}
\bar{p}_{(n)}=\frac{2\pi}{\epsilon \Omega^2 v_{(n)}}A_{(n)}.
\end{equation}
Now, the ambient field experienced by the $n$th resonator is the superposition of the incident and reflected fields, in addition to the spherical waves diffracted by all of the other resonators, evaluated at $\bx^+=\br_{(n)}$. Thus, generalizing the argument in \S\S\ref{ssec:arbinc}, the diffraction amplitudes $A_{(n)}$ satisfy the following set of $N$ algebraic equations
\begin{equation}\label{foldy}
 A_{(n)} =\left( p_{i}(\mathbf{r}_{(n)}) + p_{r}(\mathbf{r}_{(n)}) +
 \sum_{m\neq n} A_{(m)}\frac{\exp{(i \epsilon^{1/2} \Omega |\mathbf{r}_{(m)}- \mathbf{r}_{(n)}|)}}{|\mathbf{r}_{(m)}- \mathbf{r}_{(n)}|} \right) f_{(n)}, \quad n=1,2,\ldots,N,
\end{equation}
where $f_{(n)}$ is the scattering factor of the $n$th resonator
\begin{equation}\label{fn}
f_{(n)}= -\epsilon\frac{ \beta_{(n)} a_{(n)} \Omega^2} {2\pi}\left[\Omega^2 - \bar{\Omega}_{(n)}^2   + i\left(\frac{\epsilon^{3/2}}{2\pi}\frac{\beta_{(n)}^{5/2} a_{(n)}^{5/2}}{v_{(n)}^{3/2}}+ \frac{\delta}{\epsilon}\frac{\beta_{(n)}^{7/4}\theta_{(n)}}{2^{1/2} a_{(n)}^{1/4}v_{(n)}^{3/4}}\right)\right]^{-1}
\end{equation}
and $\bar{\Omega}_{(n)}$ is the dimensionless resonance frequency of the $n$th resonator 
\begin{equation}\label{Omega n}
\bar{\Omega}_{(n)}  = \left(\frac{\beta_{(n)} a_{(n)}}{v_{(n)}}\right)^{1/2} + \epsilon\frac{\beta_{(n)}^{3/2} a_{(n)}^{3/2}}{v_{(n)}^{5/6}}
\frac{\sigma_{(n)} }{2}- \frac{\delta}{\epsilon}\frac{\beta_{(n)}^{5/4}\theta_{(n)}}{2^{3/2} a_{(n)}^{3/4}v_{(n)}^{1/4}},  
\end{equation}
with $\beta_{(n)}  = \beta(h_{(n)})$, $\theta_{(n)}  = \theta(h_{(n)})$, and $\sigma_{(n)}$ is the shape factor corresponding to the cavity shape of the $n$th resonator with volume normalized to unity. 

The above multiple-resonator model is provided here as an intuitive generalization of the asymptotic single-resonator model. Our ``derivation'' of this scheme follows the paradigm of Foldy's method, a very common multiple-scattering method for calculating the response of an arrangement of strongly interacting subwavelength scatterers, most often of the monopole type \cite{carstensen:1947,martin:2006}. In Foldy's method, the scattering factor describes the frequency response of a resonator in isolation; our asymptotic single-resonator theory provides explicit formulas for the scattering factors, with losses included based on first principles. Note that Foldy's method assumes that the spacing between the scatterers is large compared to their size; fortunately, for surface-embedded Helmholtz resonators 
the pertinent scatterer size is actually the $O(\epsilon)$ neck opening, rather than the $O(1)$ linear dimension of the cavity. Thus, we can expect the model to be applicable even to the case of adjacent cavities.

\subsection{Critical coupling and perfect absorption}
\label{ssec:pabs}
As an example, we consider a metasurface formed of a doubly periodic array of identical Helmholtz resonators that is subjected to a normally incident plane wave. We assume that the centers of the neck openings are located at 
\begin{equation}\label{center position}
\mathbf{r}_{(n_{1},n_{2})}= n_{1} \boldsymbol{b}_{1} + n_{2} \boldsymbol{b}_{2},
\end{equation}
where $\boldsymbol{b}_{1}$ and $\boldsymbol{b}_{2}$ are lattice basis vectors in the plane of the substrate and $\{n_{1},n_{2}\} \in \mathbb{Z}^{2}$. The dimensionless unit-cell area is denoted by $\mathcal{A}$. We shall focus on the typical metasurface scenario where the array spacing is subwavelength, with $\mathcal{A}$ fixed as $\epsilon\to0$. 

The symmetry of the problem dictates that the diffraction amplitudes are identical, say equal to $A$. Thus inversion of \eqref{foldy} gives the formal solution
\begin{equation}\label{w2}
A =   \frac{2 f}{1 - \chi f}, \quad \chi = \sum_{(n_1,n_2)}^{'} \frac{\exp{\left(\epsilon^{1/2} \Omega| \mathbf{r}_{(n_{1},n_{2})}|\right)}}{|\mathbf{r}_{(n_{1},n_{2})}|},
\end{equation}
the dash indicating that the zeroth term should be omitted. Conventional methods can be used to transform the conditionally convergent sum into an absolutely convergent one \cite{enoch:2001,linton:2010}. As $\epsilon\to0$ with $\mathcal{A}$ fixed, that absolute convergent sum possesses the expansion \cite{linton:2010}
\begin{equation}\label{chiasym}
\chi =  i \epsilon^{-1/2}\frac{ 2 \pi}{ \Omega \mathcal{A}} + o(\epsilon^{-1/2}).
\end{equation}

Substituting \eqref{chiasym} into \eqref{w2} and discarding negligible terms yields 
\begin{equation}\label{A PA}
A =
-\frac{\epsilon}{\pi} \frac{\beta\Omega^2}{\Omega^2 - \beta\left(1 -  \frac{1}{\epsilon}\frac{\delta\beta^{5/4}\theta}{2^{3/2}}\right)^2  + i\left(\epsilon^{1/2} \frac{ \beta^{3/2} }{\mathcal{A}} +  \frac{1}{\epsilon}\frac{\delta\beta^{7/4}\theta}{2^{1/2}}\right) }.
\end{equation}
Note that the term in the denominator corresponding to radiation damping is enhanced by an $O(1/\epsilon)$ factor in comparison with the single-resonator case \eqref{uniform approx}. 
As a consequence, at resonance, $A$ is at most $O(\epsilon^{1/2})$ and $\bar{p}$ is at most $O(\epsilon^{-1/2})$; furthermore, the width of the resonance is at least $O(\epsilon^{1/2})$. In particular, the latter scaling explains the absence in \eqref{A PA} of the $O(\epsilon)$ frequency shift associated with the cavity shape factor. Since the scaling of the term associated with dissipation remains the same as before, the regime where radiation damping is comparable with dissipation loss becomes $\delta=O(\epsilon^{3/2})$; in that sense, the metasurface is less sensitive to losses than a single resonator. 

The total pressure field (at distances $\gg\epsilon$ away from the surface) can be found by substituting \eqref{A PA} into \eqref{mult pres}, with the ambient field $p_i(\bx^+)+p_r(\bx^+)$ being the same as in \eqref{farfie}. 
In particular, the far-field can then be obtained following the analysis in \cite{linton:2010}:
\begin{equation}\label{PA farfield}
p \sim \exp\left(-i \epsilon^{1/2} \Omega\, \unit\bcdot \bx^+ \right) + R\exp\left(i \epsilon^{1/2} \Omega\, \unit\bcdot \bx^+ \right) \quad \text{as} \quad  \unit\bcdot \bx^+ \to \infty,
\end{equation}
wherein 
\begin{equation}\label{R rel}
R = 1 + i \epsilon^{-1/2}\frac{2\pi A}{\Omega\mathcal{A}}
\end{equation}
constitues an effective reflection coefficient. 
Given the subwavelength spacing between the resonators, the far field contains a single reflected wave. This reflected wave, in turn, is a superposition of the wave reflected from the rigid substrate and that generated by summation of the diffracted spherical waves. 

More explicitly, substituting \eqref{A PA} into \eqref{R rel} yields 
\begin{equation}\label{R}
R = \frac{\Omega^2 - \beta\left(1 -  \frac{1}{\epsilon}\frac{\delta\beta^{5/4}\theta}{2^{3/2}}\right)^2   - i\left(\epsilon^{1/2} \frac{ \beta^{3/2} }{\mathcal{A}} - \frac{1}{\epsilon}\frac{\delta\beta^{7/4}\theta}{2^{1/2}}\right)}{\Omega^2 - \beta\left(1 -  \frac{1}{\epsilon}\frac{\delta\beta^{5/4}\theta}{2^{3/2}}\right)^2  + i\left(\epsilon^{1/2} \frac{ \beta^{3/2} }{\mathcal{A}} +  \frac{1}{\epsilon}\frac{\delta\beta^{7/4}\theta}{2^{1/2}}\right) }.
\end{equation}
From this expression is is clear that $|R|$ attains its minimum value at resonance. In particular, if the parametric relation 
\begin{equation}\label{ccoupling}
\mathcal{A}\frac{\delta}{\epsilon^{3/2}}=\frac{ 2^{1/2}}{\beta^{1/4}\theta } 
\end{equation}
is satisfied, then $R = 0$ at resonance. This particular situation corresponds to the phenomenon of perfect absorption \cite{jimenez:2016}, where the metasurface absorbs all of the incident power. 

Note that perfect absorption occurs when the terms in \eqref{R} associated with radiation damping and thermoviscous dissipation are equal. Accordingly, \eqref{ccoupling} is also the condition for critical coupling of a subwavelength metasurface. It differs from condition \eqref{critc single}, derived for a single resonator, in the scaling of $\delta$ with $\epsilon$, in the dependence upon $\mathcal{A}$, and in the function of the neck aspect ratio $h$ on the right hand side, which can again be easily calculated using Table \ref{table:parameters}. The critical-coupling condition \eqref{ccoupling} is depicted in Fig.~\ref{fig:critc}(b), for several values of the unit-cell area $\mathcal{A}$. Similar to the case of a single resonator, critical coupling for a metasurface can in principle be achieved for more than one value of the neck aspect ratio $h$ (with all other parameters the same), or not at all when $\delta/\epsilon^{3/2}$ exceeds a critical value.
 
\section{Concluding remarks}\label{sec:conclusions}

This paper provides an asymptotic model for the linear response of a surface-embedded Helmholtz resonator including thermoviscous effects, with the geometric parameters in the model precisely defined through canonical problems and provided explicitly for a wide range of geometries. 
Using Foldy's method, we extended this model to allow for multiple surface-embedded resonators, arbitrarily distributed and not necessarily identical, subjected to an arbitrary incident field. Many other straightforward extensions of the model are possible, e.g., non-cylindrical necks, multiple-neck resonators, resonators in free space, resonators connected to channels \cite{fang:2006} and resonator networks \cite{vanel:2019}. In these scenarios, thermoviscous effects could be systematically modeled along the lines of the present paper. 

Looking forward, we envisage the multiple-resonator extension of our model being used as a versatile and computationally light test-bed for studying the influence of thermoviscous losses on acoustic metasurface phenomena. It is worth emphasizing the levels of complexity that can be simulated. First, the spacing and geometrical arrangement of the resonators can be chosen at will. Second, there are four geometric parameters to tune for each non-identical resonator: neck-to-cavity ratio, neck aspect ratio, cavity shape factor and   dimensionless cavity volume (choosing to operate close to the Helmholtz resonance fixes the latter parameter for one  resonator). Lastly, the ratio between any representative length scale and the characteristic viscous length provides one additional ``global'' parameter determining the relative importance of thermoviscous effects.

Given that sound attenuation remains the primary application for Helmholtz resonators, we here chose to illustrate our model by deriving explicit critical-coupling conditions for optimal and perfect absorption of isolated resonators and metasurfaces, respectively. In both problems, critical coupling can in principle be achieved for two different values of the neck aspect ratio (with all other parameters the same); this observation builds on the result in \cite{brandao:20xx}, that the acoustic resistance of a cylindrical neck exhibits a minimum as a function of the neck aspect ratio. In practice, usually broadband absorption rather than optimal absorption at a single frequency is desirable. Our results show that that the bandwidth of a critically coupled metasurface, in the case where the spacing between the resonators is subwavelength, is asymptotically larger than the bandwidth of a critically coupled isolated resonator; this is owing to the comparable enhancement of  radiation damping in the former case (and hence also dissipation required for critical coupling). We anticipate that our multiple-resonator model could be used to design more sophisticated broadband absorbing surfaces. 

\textbf{Acknowledgements.}
OS is grateful for support from EPSRC through grant EP/R041458/1.

\bibliography{refs}

%merlin.mbs apsrev4-1.bst 2010-07-25 4.21a (PWD, AO, DPC) hacked
%Control: key (0)
%Control: author (8) initials jnrlst
%Control: editor formatted (1) identically to author
%Control: production of article title (-1) disabled
%Control: page (0) single
%Control: year (1) truncated
%Control: production of eprint (0) enabled
\begin{thebibliography}{43}%
\makeatletter
\providecommand \@ifxundefined [1]{%
 \@ifx{#1\undefined}
}%
\providecommand \@ifnum [1]{%
 \ifnum #1\expandafter \@firstoftwo
 \else \expandafter \@secondoftwo
 \fi
}%
\providecommand \@ifx [1]{%
 \ifx #1\expandafter \@firstoftwo
 \else \expandafter \@secondoftwo
 \fi
}%
\providecommand \natexlab [1]{#1}%
\providecommand \enquote  [1]{``#1''}%
\providecommand \bibnamefont  [1]{#1}%
\providecommand \bibfnamefont [1]{#1}%
\providecommand \citenamefont [1]{#1}%
\providecommand \href@noop [0]{\@secondoftwo}%
\providecommand \href [0]{\begingroup \@sanitize@url \@href}%
\providecommand \@href[1]{\@@startlink{#1}\@@href}%
\providecommand \@@href[1]{\endgroup#1\@@endlink}%
\providecommand \@sanitize@url [0]{\catcode `\\12\catcode `\$12\catcode
  `\&12\catcode `\#12\catcode `\^12\catcode `\_12\catcode `\%12\relax}%
\providecommand \@@startlink[1]{}%
\providecommand \@@endlink[0]{}%
\providecommand \url  [0]{\begingroup\@sanitize@url \@url }%
\providecommand \@url [1]{\endgroup\@href {#1}{\urlprefix }}%
\providecommand \urlprefix  [0]{URL }%
\providecommand \Eprint [0]{\href }%
\providecommand \doibase [0]{http://dx.doi.org/}%
\providecommand \selectlanguage [0]{\@gobble}%
\providecommand \bibinfo  [0]{\@secondoftwo}%
\providecommand \bibfield  [0]{\@secondoftwo}%
\providecommand \translation [1]{[#1]}%
\providecommand \BibitemOpen [0]{}%
\providecommand \bibitemStop [0]{}%
\providecommand \bibitemNoStop [0]{.\EOS\space}%
\providecommand \EOS [0]{\spacefactor3000\relax}%
\providecommand \BibitemShut  [1]{\csname bibitem#1\endcsname}%
\let\auto@bib@innerbib\@empty
%</preamble>
\bibitem [{\citenamefont {Helmholtz}(1860)}]{helmholtz:1860}%
  \BibitemOpen
  \bibfield  {author} {\bibinfo {author} {\bibfnamefont {H.~V.}\ \bibnamefont
  {Helmholtz}},\ }\href@noop {} {\bibfield  {journal} {\bibinfo  {journal}
  {Crelle}\ }\textbf {\bibinfo {volume} {57}},\ \bibinfo {pages} {1} (\bibinfo
  {year} {1860})}\BibitemShut {NoStop}%
\bibitem [{\citenamefont {Lord{ }Rayleigh}(1871)}]{rayleigh:1871}%
  \BibitemOpen
  \bibfield  {author} {\bibinfo {author} {\bibnamefont {Lord{ }Rayleigh}},\
  }\href@noop {} {\bibfield  {journal} {\bibinfo  {journal} {Phil. Trans. R.
  Soc. Lond.}\ }\textbf {\bibinfo {volume} {161}},\ \bibinfo {pages} {77}
  (\bibinfo {year} {1871})}\BibitemShut {NoStop}%
\bibitem [{\citenamefont {Lord{ }Rayleigh}(1896)}]{rayleigh:1896}%
  \BibitemOpen
  \bibfield  {author} {\bibinfo {author} {\bibnamefont {Lord{ }Rayleigh}},\
  }\href@noop {} {\emph {\bibinfo {title} {The Theory of Sound}}}\ (\bibinfo
  {publisher} {London:Macmillan},\ \bibinfo {year} {1896})\BibitemShut
  {NoStop}%
\bibitem [{\citenamefont {Ingard}(1953)}]{ingard:1953}%
  \BibitemOpen
  \bibfield  {author} {\bibinfo {author} {\bibfnamefont {U.}~\bibnamefont
  {Ingard}},\ }\href@noop {} {\bibfield  {journal} {\bibinfo  {journal} {J.
  Acoust. Soc. Am.}\ }\textbf {\bibinfo {volume} {25}},\ \bibinfo {pages}
  {1037} (\bibinfo {year} {1953})}\BibitemShut {NoStop}%
\bibitem [{\citenamefont {Maa}(1998)}]{maa:1998}%
  \BibitemOpen
  \bibfield  {author} {\bibinfo {author} {\bibfnamefont {D.~Y.}\ \bibnamefont
  {Maa}},\ }\href@noop {} {\bibfield  {journal} {\bibinfo  {journal} {J.
  Acoust. Soc. Am.}\ }\textbf {\bibinfo {volume} {104}},\ \bibinfo {pages}
  {2861} (\bibinfo {year} {1998})}\BibitemShut {NoStop}%
\bibitem [{\citenamefont {Yasuda}\ \emph {et~al.}(2013)\citenamefont {Yasuda},
  \citenamefont {Wu}, \citenamefont {Nakagawa},\ and\ \citenamefont
  {Nagamura}}]{yasuda:2013}%
  \BibitemOpen
  \bibfield  {author} {\bibinfo {author} {\bibfnamefont {T.}~\bibnamefont
  {Yasuda}}, \bibinfo {author} {\bibfnamefont {C.}~\bibnamefont {Wu}}, \bibinfo
  {author} {\bibfnamefont {N.}~\bibnamefont {Nakagawa}}, \ and\ \bibinfo
  {author} {\bibfnamefont {K.}~\bibnamefont {Nagamura}},\ }\href@noop {}
  {\bibfield  {journal} {\bibinfo  {journal} {Appl. Acoust.}\ }\textbf
  {\bibinfo {volume} {74}},\ \bibinfo {pages} {49} (\bibinfo {year}
  {2013})}\BibitemShut {NoStop}%
\bibitem [{\citenamefont {Zhao}\ and\ \citenamefont {Li}(2015)}]{zhao:2015}%
  \BibitemOpen
  \bibfield  {author} {\bibinfo {author} {\bibfnamefont {D.}~\bibnamefont
  {Zhao}}\ and\ \bibinfo {author} {\bibfnamefont {X.}~\bibnamefont {Li}},\
  }\href@noop {} {\bibfield  {journal} {\bibinfo  {journal} {Prog. Aerosp.
  Sci.}\ }\textbf {\bibinfo {volume} {74}},\ \bibinfo {pages} {114} (\bibinfo
  {year} {2015})}\BibitemShut {NoStop}%
\bibitem [{\citenamefont {Fang}\ \emph {et~al.}(2006)\citenamefont {Fang},
  \citenamefont {Xi}, \citenamefont {Xu}, \citenamefont {Ambati}, \citenamefont
  {Srituravanich}, \citenamefont {Sun},\ and\ \citenamefont
  {Zhang}}]{fang:2006}%
  \BibitemOpen
  \bibfield  {author} {\bibinfo {author} {\bibfnamefont {N.}~\bibnamefont
  {Fang}}, \bibinfo {author} {\bibfnamefont {D.}~\bibnamefont {Xi}}, \bibinfo
  {author} {\bibfnamefont {J.}~\bibnamefont {Xu}}, \bibinfo {author}
  {\bibfnamefont {M.}~\bibnamefont {Ambati}}, \bibinfo {author} {\bibfnamefont
  {W.}~\bibnamefont {Srituravanich}}, \bibinfo {author} {\bibfnamefont
  {C.}~\bibnamefont {Sun}}, \ and\ \bibinfo {author} {\bibfnamefont
  {X.}~\bibnamefont {Zhang}},\ }\href@noop {} {\bibfield  {journal} {\bibinfo
  {journal} {Nat. Mater.}\ }\textbf {\bibinfo {volume} {5}},\ \bibinfo {pages}
  {452} (\bibinfo {year} {2006})}\BibitemShut {NoStop}%
\bibitem [{\citenamefont {Lemoult}\ \emph {et~al.}(2011)\citenamefont
  {Lemoult}, \citenamefont {Fink},\ and\ \citenamefont
  {Lerosey}}]{lemoult:2011}%
  \BibitemOpen
  \bibfield  {author} {\bibinfo {author} {\bibfnamefont {F.}~\bibnamefont
  {Lemoult}}, \bibinfo {author} {\bibfnamefont {M.}~\bibnamefont {Fink}}, \
  and\ \bibinfo {author} {\bibfnamefont {G.}~\bibnamefont {Lerosey}},\
  }\href@noop {} {\bibfield  {journal} {\bibinfo  {journal} {Phys. Rev. Lett.}\
  }\textbf {\bibinfo {volume} {107}},\ \bibinfo {pages} {064301} (\bibinfo
  {year} {2011})}\BibitemShut {NoStop}%
\bibitem [{\citenamefont {Jim{\'e}nez}\ \emph {et~al.}(2016)\citenamefont
  {Jim{\'e}nez}, \citenamefont {Huang}, \citenamefont {Romero-Garc{\'\i}a},
  \citenamefont {Pagneux},\ and\ \citenamefont {Groby}}]{jimenez:2016}%
  \BibitemOpen
  \bibfield  {author} {\bibinfo {author} {\bibfnamefont {N.}~\bibnamefont
  {Jim{\'e}nez}}, \bibinfo {author} {\bibfnamefont {W.}~\bibnamefont {Huang}},
  \bibinfo {author} {\bibfnamefont {V.}~\bibnamefont {Romero-Garc{\'\i}a}},
  \bibinfo {author} {\bibfnamefont {V.}~\bibnamefont {Pagneux}}, \ and\
  \bibinfo {author} {\bibfnamefont {J.-P.}\ \bibnamefont {Groby}},\ }\href@noop
  {} {\bibfield  {journal} {\bibinfo  {journal} {Appl. Phys. Lett.}\ }\textbf
  {\bibinfo {volume} {109}},\ \bibinfo {pages} {121902} (\bibinfo {year}
  {2016})}\BibitemShut {NoStop}%
\bibitem [{\citenamefont {Ding}\ \emph {et~al.}(2016)\citenamefont {Ding},
  \citenamefont {Ma}, \citenamefont {Xiao}, \citenamefont {Zhang},\ and\
  \citenamefont {Chan}}]{ding:2016}%
  \BibitemOpen
  \bibfield  {author} {\bibinfo {author} {\bibfnamefont {K.}~\bibnamefont
  {Ding}}, \bibinfo {author} {\bibfnamefont {G.}~\bibnamefont {Ma}}, \bibinfo
  {author} {\bibfnamefont {M.}~\bibnamefont {Xiao}}, \bibinfo {author}
  {\bibfnamefont {Z.~Q.}\ \bibnamefont {Zhang}}, \ and\ \bibinfo {author}
  {\bibfnamefont {C.~T.}\ \bibnamefont {Chan}},\ }\href@noop {} {\bibfield
  {journal} {\bibinfo  {journal} {Phys. Rev. X}\ }\textbf {\bibinfo {volume}
  {6}},\ \bibinfo {pages} {021007} (\bibinfo {year} {2016})}\BibitemShut
  {NoStop}%
\bibitem [{\citenamefont {Jim{\'e}nez}\ \emph {et~al.}(2017)\citenamefont
  {Jim{\'e}nez}, \citenamefont {Romero-Garc{\'\i}a}, \citenamefont {Pagneux},\
  and\ \citenamefont {Groby}}]{jimenez:2017}%
  \BibitemOpen
  \bibfield  {author} {\bibinfo {author} {\bibfnamefont {N.}~\bibnamefont
  {Jim{\'e}nez}}, \bibinfo {author} {\bibfnamefont {V.}~\bibnamefont
  {Romero-Garc{\'\i}a}}, \bibinfo {author} {\bibfnamefont {V.}~\bibnamefont
  {Pagneux}}, \ and\ \bibinfo {author} {\bibfnamefont {J.-P.}\ \bibnamefont
  {Groby}},\ }\href@noop {} {\bibfield  {journal} {\bibinfo  {journal} {Sci.
  Rep.}\ }\textbf {\bibinfo {volume} {7}},\ \bibinfo {pages} {13595} (\bibinfo
  {year} {2017})}\BibitemShut {NoStop}%
\bibitem [{\citenamefont {Ingard}\ and\ \citenamefont
  {Labate}(1950)}]{ingard:1950}%
  \BibitemOpen
  \bibfield  {author} {\bibinfo {author} {\bibfnamefont {U.}~\bibnamefont
  {Ingard}}\ and\ \bibinfo {author} {\bibfnamefont {S.}~\bibnamefont
  {Labate}},\ }\href@noop {} {\bibfield  {journal} {\bibinfo  {journal} {J.
  Acoust. Soc. Am.}\ }\textbf {\bibinfo {volume} {22}},\ \bibinfo {pages} {211}
  (\bibinfo {year} {1950})}\BibitemShut {NoStop}%
\bibitem [{\citenamefont {Zinn}(1970)}]{zinn:1970}%
  \BibitemOpen
  \bibfield  {author} {\bibinfo {author} {\bibfnamefont {B.~T.}\ \bibnamefont
  {Zinn}},\ }\href@noop {} {\bibfield  {journal} {\bibinfo  {journal} {J. Sound
  Vib.}\ }\textbf {\bibinfo {volume} {13}},\ \bibinfo {pages} {347} (\bibinfo
  {year} {1970})}\BibitemShut {NoStop}%
\bibitem [{\citenamefont {Alster}(1972)}]{alster:1972}%
  \BibitemOpen
  \bibfield  {author} {\bibinfo {author} {\bibfnamefont {M.}~\bibnamefont
  {Alster}},\ }\href@noop {} {\bibfield  {journal} {\bibinfo  {journal} {J.
  Sound Vib.}\ }\textbf {\bibinfo {volume} {24}},\ \bibinfo {pages} {63}
  (\bibinfo {year} {1972})}\BibitemShut {NoStop}%
\bibitem [{\citenamefont {Panton}\ and\ \citenamefont
  {Miller}(1975)}]{panton:1975}%
  \BibitemOpen
  \bibfield  {author} {\bibinfo {author} {\bibfnamefont {R.}~\bibnamefont
  {Panton}}\ and\ \bibinfo {author} {\bibfnamefont {J.}~\bibnamefont
  {Miller}},\ }\href@noop {} {\bibfield  {journal} {\bibinfo  {journal} {J.
  Acoust. Soc. Am.}\ }\textbf {\bibinfo {volume} {57}},\ \bibinfo {pages}
  {1533} (\bibinfo {year} {1975})}\BibitemShut {NoStop}%
\bibitem [{\citenamefont {Chanaud}(1994)}]{chanaud:1994}%
  \BibitemOpen
  \bibfield  {author} {\bibinfo {author} {\bibfnamefont {R.}~\bibnamefont
  {Chanaud}},\ }\href@noop {} {\bibfield  {journal} {\bibinfo  {journal} {J.
  Sound Vib.}\ }\textbf {\bibinfo {volume} {178}},\ \bibinfo {pages} {337}
  (\bibinfo {year} {1994})}\BibitemShut {NoStop}%
\bibitem [{\citenamefont {Hersh}\ \emph {et~al.}(2003)\citenamefont {Hersh},
  \citenamefont {Walker},\ and\ \citenamefont {Celano}}]{hersh:2003}%
  \BibitemOpen
  \bibfield  {author} {\bibinfo {author} {\bibfnamefont {A.~S.}\ \bibnamefont
  {Hersh}}, \bibinfo {author} {\bibfnamefont {B.~E.}\ \bibnamefont {Walker}}, \
  and\ \bibinfo {author} {\bibfnamefont {J.~W.}\ \bibnamefont {Celano}},\
  }\href@noop {} {\bibfield  {journal} {\bibinfo  {journal} {AIAA J.}\ }\textbf
  {\bibinfo {volume} {41}},\ \bibinfo {pages} {795} (\bibinfo {year}
  {2003})}\BibitemShut {NoStop}%
\bibitem [{\citenamefont {Howe}(1976)}]{howe:1976}%
  \BibitemOpen
  \bibfield  {author} {\bibinfo {author} {\bibfnamefont {M.~S.}\ \bibnamefont
  {Howe}},\ }\href@noop {} {\bibfield  {journal} {\bibinfo  {journal} {J. Sound
  Vib.}\ }\textbf {\bibinfo {volume} {45}},\ \bibinfo {pages} {427} (\bibinfo
  {year} {1976})}\BibitemShut {NoStop}%
\bibitem [{\citenamefont {Sugimoto}\ and\ \citenamefont
  {Horioka}(1995)}]{sugimoto:1995}%
  \BibitemOpen
  \bibfield  {author} {\bibinfo {author} {\bibfnamefont {N.}~\bibnamefont
  {Sugimoto}}\ and\ \bibinfo {author} {\bibfnamefont {T.}~\bibnamefont
  {Horioka}},\ }\href@noop {} {\bibfield  {journal} {\bibinfo  {journal} {J.
  Acoust. Soc. Am.}\ }\textbf {\bibinfo {volume} {97}},\ \bibinfo {pages}
  {1446} (\bibinfo {year} {1995})}\BibitemShut {NoStop}%
\bibitem [{\citenamefont {Komkin}\ \emph {et~al.}(2017)\citenamefont {Komkin},
  \citenamefont {Mironov},\ and\ \citenamefont {Bykov}}]{komkin:2017}%
  \BibitemOpen
  \bibfield  {author} {\bibinfo {author} {\bibfnamefont {A.~I.}\ \bibnamefont
  {Komkin}}, \bibinfo {author} {\bibfnamefont {M.~A.}\ \bibnamefont {Mironov}},
  \ and\ \bibinfo {author} {\bibfnamefont {A.~I.}\ \bibnamefont {Bykov}},\
  }\href@noop {} {\bibfield  {journal} {\bibinfo  {journal} {Acoust. Phys.}\
  }\textbf {\bibinfo {volume} {63}},\ \bibinfo {pages} {385} (\bibinfo {year}
  {2017})}\BibitemShut {NoStop}%
\bibitem [{\citenamefont {Bigg}(1982)}]{bigg:1982}%
  \BibitemOpen
  \bibfield  {author} {\bibinfo {author} {\bibfnamefont {G.~R.}\ \bibnamefont
  {Bigg}},\ }\href@noop {} {\bibfield  {journal} {\bibinfo  {journal} {J. Sound
  Vib.}\ }\textbf {\bibinfo {volume} {85}},\ \bibinfo {pages} {85} (\bibinfo
  {year} {1982})}\BibitemShut {NoStop}%
\bibitem [{\citenamefont {Monkewitz}\ and\ \citenamefont
  {Nguyen-Vo}(1985)}]{monkewitz:1985}%
  \BibitemOpen
  \bibfield  {author} {\bibinfo {author} {\bibfnamefont {P.~A.}\ \bibnamefont
  {Monkewitz}}\ and\ \bibinfo {author} {\bibfnamefont {N.-M.}\ \bibnamefont
  {Nguyen-Vo}},\ }\href@noop {} {\bibfield  {journal} {\bibinfo  {journal} {J.
  Fluid Mech.}\ }\textbf {\bibinfo {volume} {151}},\ \bibinfo {pages} {477}
  (\bibinfo {year} {1985})}\BibitemShut {NoStop}%
\bibitem [{\citenamefont {Monkewitz}(1985)}]{monkewitz2:1985}%
  \BibitemOpen
  \bibfield  {author} {\bibinfo {author} {\bibfnamefont {P.~A.}\ \bibnamefont
  {Monkewitz}},\ }\href@noop {} {\bibfield  {journal} {\bibinfo  {journal} {J.
  Fluid Mech.}\ }\textbf {\bibinfo {volume} {156}},\ \bibinfo {pages} {151}
  (\bibinfo {year} {1985})}\BibitemShut {NoStop}%
\bibitem [{\citenamefont {Mohring}(1999)}]{mohring:1999}%
  \BibitemOpen
  \bibfield  {author} {\bibinfo {author} {\bibfnamefont {J.}~\bibnamefont
  {Mohring}},\ }\href@noop {} {\bibfield  {journal} {\bibinfo  {journal} {Acta
  Acust united Ac}\ }\textbf {\bibinfo {volume} {85}},\ \bibinfo {pages} {751}
  (\bibinfo {year} {1999})}\BibitemShut {NoStop}%
\bibitem [{\citenamefont {Ammari}\ and\ \citenamefont
  {Zhang}(2015)}]{ammari:2015}%
  \BibitemOpen
  \bibfield  {author} {\bibinfo {author} {\bibfnamefont {H.}~\bibnamefont
  {Ammari}}\ and\ \bibinfo {author} {\bibfnamefont {H.}~\bibnamefont {Zhang}},\
  }\href@noop {} {\bibfield  {journal} {\bibinfo  {journal} {Commun. Math.
  Phys.}\ }\textbf {\bibinfo {volume} {337}},\ \bibinfo {pages} {379} (\bibinfo
  {year} {2015})}\BibitemShut {NoStop}%
\bibitem [{\citenamefont {Carstensen}\ and\ \citenamefont
  {Foldy}(1947)}]{carstensen:1947}%
  \BibitemOpen
  \bibfield  {author} {\bibinfo {author} {\bibfnamefont {E.~L.}\ \bibnamefont
  {Carstensen}}\ and\ \bibinfo {author} {\bibfnamefont {L.~L.}\ \bibnamefont
  {Foldy}},\ }\href@noop {} {\bibfield  {journal} {\bibinfo  {journal} {J.
  Acoust. Soc. Am.}\ }\textbf {\bibinfo {volume} {19}},\ \bibinfo {pages} {481}
  (\bibinfo {year} {1947})}\BibitemShut {NoStop}%
\bibitem [{\citenamefont {Martin}(2006)}]{martin:2006}%
  \BibitemOpen
  \bibfield  {author} {\bibinfo {author} {\bibfnamefont {P.~A.}\ \bibnamefont
  {Martin}},\ }\href@noop {} {\emph {\bibinfo {title} {Multiple Scattering:
  Interaction of Time-Harmonic Waves with N obstacles}}}\ (\bibinfo
  {publisher} {Cambridge University Press},\ \bibinfo {year}
  {2006})\BibitemShut {NoStop}%
\bibitem [{\citenamefont {Brand{\~a}o}\ and\ \citenamefont
  {Schnitzer}()}]{brandao:20xx}%
  \BibitemOpen
  \bibfield  {author} {\bibinfo {author} {\bibfnamefont {R.}~\bibnamefont
  {Brand{\~a}o}}\ and\ \bibinfo {author} {\bibfnamefont {O.}~\bibnamefont
  {Schnitzer}},\ }\href@noop {} {\bibfield  {journal} {\bibinfo  {journal}
  {submitted}\ }}\bibinfo {note} {(arXiv:2001.04904)}\BibitemShut {NoStop}%
\bibitem [{\citenamefont {Van~Dyke}(1975)}]{Van:pert}%
  \BibitemOpen
  \bibfield  {author} {\bibinfo {author} {\bibfnamefont {M.~D.}\ \bibnamefont
  {Van~Dyke}},\ }\href@noop {} {\emph {\bibinfo {title} {Perturbation Methods
  in Fluid Dynamics}}}\ (\bibinfo  {publisher} {Parabolic Press},\ \bibinfo
  {year} {1975})\BibitemShut {NoStop}%
\bibitem [{\citenamefont {Crighton}\ \emph {et~al.}(1992)\citenamefont
  {Crighton}, \citenamefont {Dowling}, \citenamefont {Ffowcs~Williams},
  \citenamefont {Heckl},\ and\ \citenamefont {Leppington}}]{crighton:1992}%
  \BibitemOpen
  \bibfield  {author} {\bibinfo {author} {\bibfnamefont {D.~G.}\ \bibnamefont
  {Crighton}}, \bibinfo {author} {\bibfnamefont {A.~P.}\ \bibnamefont
  {Dowling}}, \bibinfo {author} {\bibfnamefont {J.~E.}\ \bibnamefont
  {Ffowcs~Williams}}, \bibinfo {author} {\bibfnamefont {M.}~\bibnamefont
  {Heckl}}, \ and\ \bibinfo {author} {\bibfnamefont {F.~G.}\ \bibnamefont
  {Leppington}},\ }\href@noop {} {\emph {\bibinfo {title} {Modern Methods in
  Analytical Acoustics}}}\ (\bibinfo  {publisher} {Springer},\ \bibinfo {year}
  {1992})\BibitemShut {NoStop}%
\bibitem [{\citenamefont {Tuck}(1975)}]{tuck:1975}%
  \BibitemOpen
  \bibfield  {author} {\bibinfo {author} {\bibfnamefont {E.~O.}\ \bibnamefont
  {Tuck}},\ }\href@noop {} {\bibfield  {journal} {\bibinfo  {journal} {Adv.
  Appl. Mech.}\ }\textbf {\bibinfo {volume} {15}},\ \bibinfo {pages} {89}
  (\bibinfo {year} {1975})}\BibitemShut {NoStop}%
\bibitem [{\citenamefont {Pierce}(1989)}]{pierce:1989}%
  \BibitemOpen
  \bibfield  {author} {\bibinfo {author} {\bibfnamefont {A.~D.}\ \bibnamefont
  {Pierce}},\ }\href@noop {} {\emph {\bibinfo {title} {Acoustics: An
  Introduction to Its Physical Principles and Applications}}}\ (\bibinfo
  {publisher} {McGraw-Hill, New York},\ \bibinfo {year} {1989})\BibitemShut
  {NoStop}%
\bibitem [{\citenamefont {Van~Bladel}(2007)}]{van:2007}%
  \BibitemOpen
  \bibfield  {author} {\bibinfo {author} {\bibfnamefont {J.~G.}\ \bibnamefont
  {Van~Bladel}},\ }\href@noop {} {\emph {\bibinfo {title} {Electromagnetic
  fields}}}\ (\bibinfo  {publisher} {John Wiley \& Sons},\ \bibinfo {year}
  {2007})\BibitemShut {NoStop}%
\bibitem [{\citenamefont {Jackson}(1999)}]{jackson:book}%
  \BibitemOpen
  \bibfield  {author} {\bibinfo {author} {\bibfnamefont {J.~D.}\ \bibnamefont
  {Jackson}},\ }\href@noop {} {\emph {\bibinfo {title} {Classical
  electrodynamics}}}\ (\bibinfo  {publisher} {American Association of Physics
  Teachers},\ \bibinfo {year} {1999})\BibitemShut {NoStop}%
\bibitem [{\citenamefont {Enoch}\ \emph {et~al.}(2001)\citenamefont {Enoch},
  \citenamefont {McPhedran}, \citenamefont {Nicorovici}, \citenamefont
  {Botten},\ and\ \citenamefont {Nixon}}]{enoch:2001}%
  \BibitemOpen
  \bibfield  {author} {\bibinfo {author} {\bibfnamefont {S.}~\bibnamefont
  {Enoch}}, \bibinfo {author} {\bibfnamefont {R.~C.}\ \bibnamefont
  {McPhedran}}, \bibinfo {author} {\bibfnamefont {N.~A.}\ \bibnamefont
  {Nicorovici}}, \bibinfo {author} {\bibfnamefont {L.~C.}\ \bibnamefont
  {Botten}}, \ and\ \bibinfo {author} {\bibfnamefont {J.~N.}\ \bibnamefont
  {Nixon}},\ }\href@noop {} {\bibfield  {journal} {\bibinfo  {journal} {J.
  Math. Phys.}\ }\textbf {\bibinfo {volume} {42}},\ \bibinfo {pages} {5859}
  (\bibinfo {year} {2001})}\BibitemShut {NoStop}%
\bibitem [{\citenamefont {Linton}(2010)}]{linton:2010}%
  \BibitemOpen
  \bibfield  {author} {\bibinfo {author} {\bibfnamefont {C.~M.}\ \bibnamefont
  {Linton}},\ }\href@noop {} {\bibfield  {journal} {\bibinfo  {journal} {SIAM
  Rev.}\ }\textbf {\bibinfo {volume} {52}},\ \bibinfo {pages} {630} (\bibinfo
  {year} {2010})}\BibitemShut {NoStop}%
\bibitem [{\citenamefont {Vanel}\ \emph {et~al.}(2019)\citenamefont {Vanel},
  \citenamefont {Craster},\ and\ \citenamefont {Schnitzer}}]{vanel:2019}%
  \BibitemOpen
  \bibfield  {author} {\bibinfo {author} {\bibfnamefont {A.~L.}\ \bibnamefont
  {Vanel}}, \bibinfo {author} {\bibfnamefont {R.~V.}\ \bibnamefont {Craster}},
  \ and\ \bibinfo {author} {\bibfnamefont {O.}~\bibnamefont {Schnitzer}},\
  }\href@noop {} {\bibfield  {journal} {\bibinfo  {journal} {SIAM J. Appl.
  Math}\ }\textbf {\bibinfo {volume} {79}},\ \bibinfo {pages} {506} (\bibinfo
  {year} {2019})}\BibitemShut {NoStop}%
\bibitem [{\citenamefont {Nijboer}\ and\ \citenamefont
  {Wette}(1957)}]{nijboer:1957}%
  \BibitemOpen
  \bibfield  {author} {\bibinfo {author} {\bibfnamefont {B.~R.~A.}\
  \bibnamefont {Nijboer}}\ and\ \bibinfo {author} {\bibfnamefont {F.~W.~D.}\
  \bibnamefont {Wette}},\ }\href@noop {} {\bibfield  {journal} {\bibinfo
  {journal} {Physica}\ }\textbf {\bibinfo {volume} {23}},\ \bibinfo {pages}
  {309} (\bibinfo {year} {1957})}\BibitemShut {NoStop}%
\bibitem [{\citenamefont {Gradshteyn}\ and\ \citenamefont
  {Ryzhik}(2014)}]{gradshteyn:2014}%
  \BibitemOpen
  \bibfield  {author} {\bibinfo {author} {\bibfnamefont {I.~S.}\ \bibnamefont
  {Gradshteyn}}\ and\ \bibinfo {author} {\bibfnamefont {I.~M.}\ \bibnamefont
  {Ryzhik}},\ }\href@noop {} {\emph {\bibinfo {title} {Table of Integrals,
  Series, and Products}}}\ (\bibinfo  {publisher} {Academic Press},\ \bibinfo
  {year} {2014})\BibitemShut {NoStop}%
\bibitem [{\citenamefont {Landau}\ and\ \citenamefont
  {Lifshitz}(1987)}]{landau:1987}%
  \BibitemOpen
  \bibfield  {author} {\bibinfo {author} {\bibfnamefont {L.~D.}\ \bibnamefont
  {Landau}}\ and\ \bibinfo {author} {\bibfnamefont {E.~M.}\ \bibnamefont
  {Lifshitz}},\ }\href@noop {} {\emph {\bibinfo {title} {Fluid Mechanics,
  Second Edition: Volume 6 (Course of Theoretical Physics)}}}\ (\bibinfo
  {publisher} {Butterworth-Heinemann},\ \bibinfo {year} {1987})\BibitemShut
  {NoStop}%
\bibitem [{\citenamefont {Rossing}(2007)}]{rossing:2007}%
  \BibitemOpen
  \bibfield  {author} {\bibinfo {author} {\bibfnamefont {T.}~\bibnamefont
  {Rossing}},\ }\href@noop {} {\emph {\bibinfo {title} {Springer Handbook of
  Acoustics}}}\ (\bibinfo  {publisher} {Springer},\ \bibinfo {year}
  {2007})\BibitemShut {NoStop}%
\bibitem [{\citenamefont {Born}\ and\ \citenamefont {Wolf}(1999)}]{born:1999}%
  \BibitemOpen
  \bibfield  {author} {\bibinfo {author} {\bibfnamefont {M.}~\bibnamefont
  {Born}}\ and\ \bibinfo {author} {\bibfnamefont {E.}~\bibnamefont {Wolf}},\
  }\href@noop {} {\emph {\bibinfo {title} {Principles of Optics:
  Electromagnetic Theory of Propagation, Interference and Diffraction of
  Light}}},\ \bibinfo {edition} {7th}\ ed.\ (\bibinfo  {publisher} {Cambridge
  University Press},\ \bibinfo {year} {1999})\BibitemShut {NoStop}%
\end{thebibliography}%

\appendix

\section{Cavity shape factor}\label{sec:csf}

\subsection{Hemisphere}
We begin by reviewing the calculation of the shape factor for a hemispherical cavity \cite{bigg:1982}. We consider the cavity to be formed of a hemisphere centered about $\bx^-=\bzero$, at which point the function $g$ is singular. In line with our normalization scheme, the volume of the hemisphere is unity and the flat face of the domain coincides with the plane $\unit\bcdot\bx^-=0$. For this geometry, the solution of the cavity problem \eqref{g eq}--\eqref{g mean} is readily found as
\begin{equation}
g  = -\frac{|\bx^{-}|^2}{6} - \frac{1}{2\pi |\bx^{-}|} + \frac{9}{10\pi}\left(\frac{2\pi}{3}\right)^{1/3}.
\end{equation}
From this solution, definition \eqref{sigma def}  yields the shape factor
\begin{equation}
\sigma_{\text{hemisphere}} = \frac{9}{10\pi}\left(\frac{2\pi}{3}\right)^{1/3}.
\end{equation}

\subsection{Cube} 
Consider now the shape factor for a cubic cavity with the singular point at the center of the face coinciding with the plane $\unit\bcdot\bx^-=0$. The symmetry of the cavity together with the Neumann condition \eqref{g neu} enables us to evenly extend $g$ about the latter plane; the geometry of the problem then consists of a rectangular domain (width $2$, unity cross-sectional area). Furthermore, the even extension of $g$ can also be periodically extended about the three symmetry planes of the rectangular domain. 

In light of the above symmetries,  the solution of the canonical cavity problem for a cube can be directly related to the lattice Green's function for an array of identical point singularities of magnitude $2$, located at the lattice position vectors
\begin{equation}
\mathbf{d} = 2 d_{1} \unit + d_{2} \boldsymbol{\hat{\jmath}} + d_{3} \boldsymbol{\hat{k}},
\end{equation}
where $\unit$, $\boldsymbol{\hat{\jmath}}$ and $\boldsymbol{\hat{k}}$ form an orthonormal basis and $\{d_{1},d_{2},d_{3}\} \in \mathbb{Z}^{3}$. A formal expression for that Green's function, which is consistent with \eqref{g eq}--\eqref{g mean}, is given by \cite{nijboer:1957,vanel:2019}
\begin{equation}\label{g cube}
g = -\frac{1}{4 \pi^2} \sum_{\mathbf{m}}^{'}\frac{e^{i 2 \pi \mathbf{m}\cdot \bx^{-} }}{|\mathbf{m}|^2},
\end{equation}
where the sum is over the set of reciprocal lattice vectors
\begin{equation}
\mathbf{m} = \frac{1}{2}m_{1} \unit + m_{2} \boldsymbol{\hat{\jmath}} + m_{3} \boldsymbol{\hat{k}}
\end{equation}
with $\{m_{1},m_{2},m_{3}\} \in \mathbb{Z}^{3}$; the dash indicates that the term corresponding to $\mathbf{m} = \mathbf{0}$ should be omitted from the summation. 

Although the series appearing in \eqref{g cube} is conditionally convergent, it can be transformed into an absolutely convergent series  \cite{nijboer:1957}:
\begin{equation}\label{g cube 2}
g =  \frac{1}{4 \pi}-\frac{1}{4 \pi^2} \sum_{\mathbf{m}}^{'}\frac{e^{-
\pi|\mathbf{m}|^2} }{|\mathbf{m}|^2}e^{i 2 \pi \mathbf{m}\cdot \bx^{-} }  -  \frac{1}{2\pi^{3/2}} \sum_{\mathbf{d}}\frac{\Gamma\left(1/2, \pi |\bx^{-} - \mathbf{d}|^2\right) }{| \bx^{-} - \mathbf{d}|},
\end{equation}
where $\Gamma\left(n, x\right)$ is the incomplete Gamma function \cite{gradshteyn:2014}. From definition \eqref{sigma def}, we find the following expression for the shape factor of a cubic cavity:
\begin{equation}
\sigma_{cube} = \frac{5}{4 \pi} -\frac{1}{4 \pi^2} \sum_{\mathbf{m}}^{'}\frac{e^{-\pi|\mathbf{m}|^2} }{|\mathbf{m}|^2} -  \frac{1}{2\pi^{3/2}} \sum_{\mathbf{d}}^{'}\frac{\Gamma\left(1/2, \pi |\mathbf{d}|^2\right) }{|\mathbf{d}|}.
\end{equation}
Evaluating the sums yields $\sigma_{\text{cube}} \approx 0.2874$.

We note that an analogous strategy could be used to compute $\sigma$ for arbitrary rectangular cavities.

\subsection{Cylinder}
Consider next a cylindrical cavity whose singular point is at the center of the base coinciding with the plane $\unit\bcdot\bx^-=0$. It is convenient to define  cylindrical coordinates $(\varrho,\varphi,z)$, where the $z$ axis is co-linear with the cylinder's axis, in the $\unit$ direction, and centered about the singular point. We denote the cylinder's radius and height by $s$ and $2\tau s$, respectively. Since in our normalization scheme  the volume of the cavity is unity, these geometric parameters are related through the relation
\begin{equation}\label{s relation}
s = \left(\frac{1}{2\pi \tau}\right)^{1/3}.
\end{equation}

As in the cubic case, we exploit the Neumann condition \eqref{g neu} to evenly extend $g$ about the plane $z=0$; accordingly, $g$ satisfies
\begin{equation}\label{g delta}
\nabla^2 g = 2\delta(\bx^{-}) - 1
\end{equation}
for $\{\varrho<s,|z|<2\tau s\}$, to be solved together with the Neumann condition \eqref{g neu} over the boundaries of the extended cylindrical domain, and the zero-mean condition \eqref{g mean}. Note that the singularity condition \eqref{g singularity} is implied by the Dirac-delta function appearing in \eqref{g delta}.

We seek a solution in the form of a Fourier--Bessel series: 
\begin{equation}\label{g bessel}
g = \sum_{n = 0}^{\infty} g_{n}(z) J_{0}\left(\frac{\lambda_{n}}{s} \varrho \right),
\end{equation}
where $\lambda_{n}$ is the $n$th root of the first-kind Bessel function of order one, $J_{1}(\cdot)$, such that the boundary condition at $\varrho=s$ is identically satisfied.
Substituting \eqref{g bessel} into \eqref{g delta} and using the orthogonality of Bessel functions, we arrive at
\begin{equation}
\frac{d^2 g_{0}}{dz^2}  = \frac{2}{\pi s^2}\delta(z)-1
%2\frac{\delta(z)}{\pi s^2}- 1
\end{equation}
and
\begin{equation}
\frac{d^2 g_{n}}{dz^2} - \frac{\lambda_{n}^2}{s^2}  g_{n}  = 2\frac{\delta(z)}{\pi s^2 J_{0}\left(\lambda_{n} \right)^2},\qquad n\neq0.
\end{equation}
The solution for $g_0$ is 
\begin{equation}
g_{0}  = - \frac{4 s^2 \tau^2}{3} + \frac{|z|}{\pi s^2}- \frac{1}{2}z^2,
\end{equation}
where the constant term is obtained from \eqref{g mean} and \eqref{s relation}. Similarly, for $g_{n}$, we find
\begin{equation}
g_{n}  =  -\frac{\cosh\left(\frac{\lambda_{n}}{s}\left[|z| - 2 s \tau\right]\right)}{\pi s \lambda_{n} J_{0}\left(\lambda_{n} \right)^2\sinh\left(2\lambda_{n} \tau \right)}.
\end{equation}
Thus, from \eqref{g bessel}, we find the solution
\begin{equation}\label{g cyl sol}
g = - \frac{4 s^2 \tau^2}{3} + \frac{|z|}{\pi s^2}- \frac{1}{2}z^2 - \frac{1}{\pi s}\sum_{n = 1}^{\infty}  \frac{\cosh\left(\frac{\lambda_{n}}{s}\left[|z| -  2 s \tau\right])\right)}{\lambda_{n} J_{0}\left(\lambda_{n} \right)^2\sinh\left(2\lambda_{n} \tau\right)}J_{0}\left(\frac{\lambda_{n}}{s} \varrho \right).
\end{equation}

The anticipated singularity of $g$ at $\bx^- = \mathbf{0}$ remains hidden in the Fourier--Bessel series appearing in \eqref{g cyl sol}. To calculate $\sigma$ from \eqref{sigma def}, we set $\varrho=0$ in \eqref{g cyl sol} and use the auxiliary series
\begin{equation}\label{g cyl aux}
\sum_{n = 1}^{\infty} \exp{\left(-\frac{\pi n}{s} |z|\right)} = \frac{1}{\exp{\left(\frac{\pi}{s}|z|\right)} - 1}
\end{equation}
to arrive at
\begin{multline}
\label{g cyl sol 2}
{\left.g\right|}_{\varrho=0}(z) = - \frac{4 s^2 \tau^2}{3} + \frac{|z|}{\pi s^2}- \frac{1}{2}z^2 - \frac{1}{2 s}\frac{1}{\exp{\left(\frac{\pi}{s}|z|\right)} - 1}  \nonumber \\- \frac{1}{\pi s}\sum_{n = 1}^{\infty}  \left(\frac{\cosh\left(\frac{\lambda_{n}}{s}\left[|z| - 2 s \tau\right]\right)}{\lambda_{n} J_{0}\left(\lambda_{n} \right)^2\sinh\left(2\lambda_{n} \tau\right)} - \frac{\pi}{2}\exp{\left(-\frac{\pi n}{s} |z|\right)} \right),
\end{multline}
wherein the series converges for $z=0$ (using the large-$n$ asymptotics of $J_{0}\left(\lambda_{n} \right)$, it can be shown that the summand is $O(n^{-2})$ as $n\to\infty$ \cite{gradshteyn:2014}). We thus find the asymptotic relation
\begin{equation}
{\left.g\right|}_{\varrho=0}(z) \sim \; -\frac{1}{2\pi |z|} + \frac{1}{4 s} -  \frac{4 s^2 \tau^2}{3}  - \frac{1}{\pi s}\sum_{n = 1}^{\infty}\left(  \frac{\coth\left(2 \lambda_{n} \tau \right)}{\lambda_{n} J_{0}\left(\lambda_{n} \right)^2} - \frac{\pi}{2}\right) + o(1) \quad \text{as} \quad z\to0.
\end{equation}
Thus, comparison with \eqref{sigma def} yields the shape factor
\begin{equation}
\sigma_{\text{cylinder}} = \left(2\pi\tau\right)^{1/3}\left[\frac{1}{4} -  \frac{2\tau}{3 \pi} - \frac{1}{\pi}\sum_{n = 1}^{\infty}\left(  \frac{\coth\left(2\lambda_{n}\tau\right)}{\lambda_{n} J_{0}\left(\lambda_{n} \right)^2} - \frac{\pi}{2}\right)\right],
\end{equation}
where \eqref{s relation} was used  to simplify the final expression.

\section{Scaling estimates}\label{app:energy}
\subsection{Energy and dissipation}\label{scaling}

Starting from the thermoviscous equations \eqref{ta1}--\eqref{ta3}, it is straightforward to derive the following energy-dissipation relation for a fluid volume $\mathcal{V}$ \cite{landau:1987,pierce:1989,rossing:2007}:
\begin{equation}\label{encon}
D_v + D_{t}  + \int_\mathcal{\partial V} \bn\bcdot\bJ\,dS =0.
\end{equation}
The first two terms are respectively the time-averaged rates of viscous and thermal energy dissipation (normalized by $p_{\infty}^2 l^2/\epsilon^{1/2} c \rho$), which are defined by the integrals
\begin{equation}\label{Dv}
D_{v} = \frac{\delta^2}{4} \int_\mathcal{V} \tE \boldsymbol{:}\tE^{*} dV,  \qquad  D_{t} =\epsilon \frac{\left(\gamma - 1\right)\delta^2}{2\Pr} \int_\mathcal{V} |\nabla T|^2 dV;
\end{equation}
in \eqref{Dv}, the asterisk denotes complex conjugation and the tensor $\tE$ is defined as 
\begin{equation}\label{E}
\tE = \bnabla \bv + \bnabla \bv^{\dagger} - \frac{2}{3}\left(\bnabla\cdot\bv\right)\tI,
\end{equation}
the dagger denoting transpose and $\tI$ being the identity tensor. The third term in \eqref{encon} represents energy leaving the fluid volume through its boundary $\partial\mathcal{V}$, where the flux vector $\bJ$ is defined as 
\begin{equation}\label{intensity}
 \bJ = \frac{1}{2} \operatorname{Re}\left( p \bv^{*} - \delta^2 \tE \cdot \bv^{*} - \epsilon \frac{ \delta^2 \left(\gamma - 1\right)}{\Pr} T \bnabla T^* \right).
\end{equation}

\subsection{Dissipation estimates} \label{ssec:dissest}
Building on the lossless analysis, it is possible to  derive order-of-magnitude estimates for the viscous and thermal dissipation rates defined in \eqref{Dv}. We begin with the viscous dissipation rate $D_v$ and focus on the neck and cavity regions. As in the main text, our main assumption is $\delta\ll\epsilon$. In that case, we anticipate the flow field $\bv$ to be approximately inviscid in the bulk of the fluid domain, including in the neck region. Indeed, the momentum equation \eqref{ta1} implies that viscous effects are only important in an oscillatory Stokes boundary layer of thickness $O(\delta)$. The tangential velocity in that layer is on the order of the local bulk velocity, say $U$, and rapidly attenuates across the layer in order to satisfy the no-slip condition \eqref{no slip}. The corresponding estimate for the viscous dissipation within the boundary layer is therefore $\delta U^2$ per unit surface area.

While the order of the local bulk velocity $U$ differs in the cavity and neck regions, the pressure scaling, say $P$, is the same in both regions. In fact, alluding to the lossless analysis, we see that $U=P/\epsilon$ and $U=\epsilon P$ respectively in the neck and cavity regions. Since the surface area of the neck is $O(\epsilon^2)$, the rate of viscous dissipation contributed by the boundary layer in that region is on the order of $D_v=\delta P^2$. Similarly, since the surface area of the cavity is $O(1)$, the viscous boundary layer in that region gives the estimate $D_v=\epsilon^2\delta P^2$.

We can similarly estimate the  thermal dissipation rate $D_t$. Since the Prandtl number is of order unity, the thickness of the thermal boundary layer is asymptotically comparable to the viscous one. From \eqref{ta3}, the bulk temperature scaling is on the order of the bulk pressure scaling $P$, in both the neck and cavity regions, and rapidly attenuates across the layer in order to satisfy the boundary condition \eqref{no temp}. It then follows from \eqref{Dv} that the thermal dissipation within the boundary layer is $\epsilon \delta P^2$ per unit surface area. Thus, we find the scaling estimates $D_t=\epsilon^3\delta P^2$ and $D_t=\epsilon \delta P^2$ based on the neck and cavity surface areas, respectively.

To summarize, the dissipation rate is dominated by an  $O(\delta P^2)$ contribution of the viscous boundary layer in the neck region, followed by an $O(\epsilon)$-smaller contribution of the thermal boundary layer in the cavity region. It can be readily verified that the contributions of the boundary layers in the wavelength-scale exterior region are relatively negligible.

\subsection{Approximate energy balance}\label{ssec:optical}
Consider the exact energy-dissipation relation \eqref{encon}, with $\mathcal{V}$ chosen as the fluid domain composed of the cavity, neck and the portion of the exterior region included in the ball $|\bx^+|<\lambda/\epsilon^{1/2}$, where $\lambda$ is a positive constant. Note that $\bn\bcdot\bJ$ vanishes at the rigid boundary because of boundary conditions \eqref{no slip} and \eqref{no temp}, so we only need to consider the contribution from the exterior wavelength-scale hemisphere boundary. Since $\delta\ll1$, effects of dissipation are negligible on the wavelength scale. Thus the form of the exterior field is approximately the same as in the lossless case \eqref{farfie}. Furthermore, since $\epsilon\ll1$, the diffracted field is still dominated by the outgoing spherical wave in that expansion. With these approximations, the flux integral in \eqref{encon} can be directly evaluated in terms of the diffraction amplitude $A$. This gives the approximate energy-dissipation balance
\begin{equation}\label{optical theorem}
\frac{2\pi}{\Omega} \textrm{Im}A = D_v+D_t + \pi\epsilon^{1/2}|A|^2.
\end{equation}

The left-hand side represents the extinction of the ambient field by the resonator, while the last term on the right-hand side represents radiation damping. Thus \eqref{optical theorem} represents the expected result that the energy removed by the ambient field is equal to the sum of the energy absorbed plus the energy scattered \cite{born:1999}.

\subsection{Diffraction amplitude and cavity pressure}
The approximate energy relation \eqref{optical theorem} can be used to derive scaling estimates for the diffraction amplitude and cavity pressure at resonance, a bound on the dissipation rate and even exact relations that can be used to corroborate the asymptotic formulas derived in the main text. 

To begin with, consider \eqref{optical theorem} together with the estimates $D_v+D_t \approx D_v =  O(\delta P^2)$ and $A=O(\epsilon P)$. The latter estimate generally follows from the discussion following \eqref{pressure exterior off} and it is further noted that $A$ is in phase with the cavity pressure. It follows that $D_v=O(\delta A^2/\epsilon)$ and thence that the absorption term in \eqref{optical theorem} is negligible for $\delta\ll\epsilon^{5/2}$, comparable to the scattering term for $\delta=O(\epsilon^{5/2})$ and dominates the scattering term for $\delta\gg\epsilon^{5/2}$.

To elaborate on these three regimes let $A=|A|e^{i\varphi}$ and note that \eqref{optical theorem} implies $\sin\varphi>0$. For $\delta\ll\epsilon^{5/2}$, the dominant balance between extinction and radiation damping gives 
\begin{equation}\label{scaling radiation dominated}
|A| \approx  \frac{2}{\Omega \epsilon^{1/2}}\sin\varphi=O\left(\frac{1}{\epsilon^{1/2}}\sin\varphi\right), \quad P=\frac{1}{\epsilon^{3/2}}\sin\varphi. 
\end{equation}
(Recall that $P$ stands for the order of magnitude of the cavity pressure.) The maximal scalings are obtained when the response is out of phase with the incident field. Namely, $A=O(\epsilon^{-1/2})$ and $P=\epsilon^{-3/2}$ at resonance, in agreement with the analysis of the lossless model. In fact, \eqref{scaling radiation dominated}, together with \eqref{leading resonance}, implies $A\sim 2i/\sqrt{\beta\epsilon}$ at resonance, in agreement with the on-resonance asymptotics \eqref{on results}.

In the opposite case, $\epsilon^{5/2}\ll\delta$, the dominant balance between extinction and absorption gives
\begin{equation}
|A|\approx \frac{\Omega D_v}{2\pi \sin\varphi}=O\left(\frac{\epsilon^2}{\delta}\sin\varphi\right), \quad P=\frac{\epsilon}{\delta}\sin\varphi.
\end{equation}
Now we have $A=O(\epsilon^2/\delta)$ and $P=O(\epsilon/\delta)$ at resonance. Note that our scalings are only valid in the thin-boundary-layer limit $\delta\ll\epsilon$, where these scalings suggest that the cavity pressure at resonance remains large and hence that the resonance remains weakly damped. 

Lastly, in the borderline case $\delta=O(\epsilon^{5/2})$, the dominant balance between extinction, absorption and radiation damping yields
\begin{equation}\label{borderline scaling}
|A|\approx\frac{\sin\varphi\pm\sqrt{\sin^2{\varphi}-\frac{\Omega^2}{\pi}\epsilon^{1/2}D_v}}{\Omega\epsilon^{1/2}}=O\left(\frac{1}{\epsilon^{1/2}}\sin\varphi\right), \quad P=\frac{1}{\epsilon^{3/2}}\sin\varphi.
\end{equation}
Thus the scalings are the same as in the case where radiation damping dominates absorption, though the above expression for $|A|$ shows that the maximal value of $|A|$ is diminished relative to the lossless case. The same expression also implies an approximate upper bound on $D_v$,
\begin{equation}
D_v\le  \frac{\pi}{\epsilon^{1/2}\Omega^2},
\end{equation}  
which is attained in the critical-coupling scenario where the absorption and radiation damping terms in \eqref{optical theorem} are equal.

\subsection{Boundary layer effects on the bulk pressure fields}
The above scalings, based on energy conservation, determine the orders of magnitude of the cavity pressure and diffraction amplitude, depending on the smallness of $\delta$ relative to $\epsilon$. An alternative viewpoint can be gained by estimating the scalings of the bulk pressure perturbations induced by the viscous and thermal boundary layers. As discussed in \S\S\ref{ssec:tbl}, although these perturbations are small, they play an important leading-order role if they are at least comparable to the small pressure perturbations found to be important in the lossless analysis of sections \S\ref{sec:leading}--\S\ref{sec:rdamp}.

We begin by estimating the magnitude of the bulk-pressure perturbations induced by displacement of the viscous boundary layer. Consider a generic  region of characteristic length scale $L$, with $U$ and $U'$ respectively denoting the orders of the bulk velocity and the perturbation to the bulk velocity induced by the viscous boundary layer. We can determine $U'$ as a function of $U$ by considering mass conservation in a fluid slab adjacent to the boundary, of area $L^2$ and thickness $\delta$. Thus, comparing the flux displaced by the boundary layer, $U'L^2$, and the net flux through the side faces of the slab, $\delta L U$, we find $U'=\delta L^{-1}U$.
This scaling relation can be rewritten in terms of pressures as $P'=\delta L^{-1}\Delta P$, where $\Delta P=LU$ is the order of bulk pressure variations and $P'=LU'$ is the order of the bulk pressure perturbation induced by the viscous boundary layer. Let us now apply these scalings to the neck, cavity and exterior regions. As before, we denote by $P$ the order of the bulk pressure in the cavity and neck regions; the pressure in the exterior is always of order unity. For the neck region, $L=\epsilon$ and $\Delta P = P$, thus $P' = \epsilon^{-1}\delta P$. For the cavity region, $L=1$ and $\Delta P =\epsilon P$, thus $P' =\epsilon \delta P$. In the exterior region, $L=\epsilon^{-1/2}$ and $\Delta P=1$, thus $P' = \epsilon^{1/2}\delta$.

Consider next the bulk-pressure perturbations owing to the thermal boundary layer. We again employ mass conservation in a thin slab defined as above. This time, however, we balance the displaced volume flux, of order $U'L^2$, and the integral over the slab volume of the term in the continuity equation \eqref{ta2} that is proportional to $p-T$; this term approximately vanishes in the bulk region but is of the order of the bulk pressure field within the thermal boundary layer. Since the volume of the slab is $\delta L^2$, we find the balance $U'=\epsilon \delta P$ for the neck and cavity regions and $U'=\epsilon \delta$ for the exterior region. Equivalently, we have that $P'=\epsilon^2 \delta P $, $P'=\epsilon \delta P$ and $P'=\epsilon^{1/2}\delta$ in the neck, cavity and exterior regions, respectively.  
\end{document}